\documentclass[10pt]{article}

\usepackage{amsmath}
\usepackage{amssymb}

\usepackage{graphicx}

\usepackage{cite}

\usepackage{color}

\usepackage[labelfont=bf,labelsep=period,justification=raggedright]{caption}
\usepackage[labelfont=bf,labelsep=period,justification=raggedright]{subfig}

\makeatletter
\renewcommand{\@biblabel}[1]{\quad#1.}
\makeatother

\date{}

\begin{document}

\begin{flushleft}
{\Large
\textbf{Dynamical modeling of collective behavior from pigeon flight data:  
flock cohesion and dispersion}
}
\\
Graciano Dieck Kattas$^{1,\ast}$, 
Xiao-Ke Xu$^{1,2}$, 
Michael Small$^{1}$
\\
\bf{1}Department of Electronic and Information Engineering, The Hong Kong Polytechnic University, Hung Hom, Kowloon, Hong Kong
\\
\bf{2} School of Communication and Electronic Engineering, Qingdao Technological University, Qingdao, China
\\

$\ast$ E-mail: g.dieck@polyu.edu.hk
\end{flushleft}

\section*{Abstract}

Several models of flocking have been promoted based on simulations with qualitatively naturalistic behavior. In this paper we provide the first direct application of computational modeling methods to infer flocking behavior from experimental field data. We show that this approach is able to infer general rules for interaction, or lack of interaction, among members of a flock or, more generally, any community. Using experimental field measurements of  
homing pigeons in flight we demonstrate the existence of a basic distance dependent attraction/repulsion relationship and show that this rule is sufficient to explain collective behavior observed in nature. Positional data of individuals over time are used as input data to a computational algorithm capable of building complex nonlinear functions that can represent the system behavior. Topological nearest neighbor interactions are considered to characterize the components within this  
model. The efficacy of this method is demonstrated with simulated noisy data generated from the classical (two dimensional) Vicsek model. When applied to experimental data from homing pigeon flights we show that the more complex three dimensional models are capable of predicting and simulating trajectories, as well as exhibiting realistic collective dynamics. The simulations of the reconstructed models are used to extract properties of the collective behavior
in pigeons, and how it is affected by changing the initial conditions of the system. Our results demonstrate that this approach may be applied to construct models capable of simulating trajectories and collective dynamics using  
experimental field measurements of herd movement. From these models, the behavior of the individual agents (animals) may be inferred.

\section*{Author Summary}

The construction of mathematical models from experimental time-series data has been considered with some success in many areas of science and engineering, using the power of computer algorithms to build model structures and tuning their parameters suitably. When considering complex systems with nonlinear or collective behavior, computational models built from real data are the alternative to emulate the system as best as possible, since classic modeling approaches based on observation should prove difficult to identify the complex dynamics. In this study, we provide a method to build models of collective dynamics from homing pigeon flight data. We show that our models follow the source dynamics well, and from them we are able to infer that significant collective behavior occurs in pigeon flights. Our results are consistent with the basic principles of previous hypotheses and models that have been proposed. Our approach serves as an initial outline towards the usage of experimental data to construct computational models to understand many complex phenomena with hypothesized collective behavior.

\section*{Introduction}

The collective behavior exhibited by interacting individuals in a population has recently attracted interest in scientific and engineering communities. Many different definitions have been used to formally describe this behavior, in order to establish new theories and further advance the contributions to this new field. In simple words, collective behavior can be described as local actions taken by individuals in a socially interacting group, which are directly related to the conditions of the group and that somehow affect the overall behavior of the group as a single global entity. Many kinds of systems from different areas of application are known to exhibit such behavior, ranging from areas like sociology, psychology, zoology, and all the way to more technical disciplines like bioengineering, computer science, and robotics. The objective of analyzing the collective behavior of a system can be either to further understand the system in question, or to apply the observed behavioral structures in other systems or circumstances in order to provide innovative solutions to problems.   

The movement of groups of animals is a common and well studied example involving the emergence of collective behavior in an interacting population. It is well known that animals tend to work in groups to achieve goals; simple examples that can come to mind are ant colonies, herds, fish schools, and bird flocks. In particular, the collective movement of a group of animals in the same direction is referred to as swarming. Bird flocking in particular, has attracted much recent attention. The ability to gain accurate positional data from GPS devices on pigeons, has opened the door to more advanced and meaningful analysis of flocking \cite{Nagy2010}. Photographic data has also lead to deeper analysis of the interaction properties of flocking \cite{Ballerini2008}. In general, with accurate 3D positional data, it is now possible to perform statistical analysis which leads to the understanding of the structural and behavioral properties of flocking. 

Mathematical models, especially dynamical models, have been used by biologists, physicists, and mathematicians, to illustrate animal movement or interaction, though usually the dynamics have been observed to be not suitable for linear models. The classical Lotka-Volterra equations are a good example of early attempts to model the growth and decay of populations of predators and preys over time, using nonlinear ordinary differential equations (ODEs). Later efforts to characterize social animal movement involved models that use physical laws and diffusion equations to describe the movement of groups of fish, insects, and herds \cite{Okubo2001}. Discrete-time generic models of collective systems, with no particular application, but which can be used to simulate swarming with complex behavior using very simple mathematical rules, are an initial step towards the full understanding of the nonlinear properties in real systems, or even graphic visualization. The Boids model \cite{Reynolds1987} is a well known example, and it has been used in movies and video games to generate 3D animated collective movement of animals. Another such case is the Vicsek model \cite{Vicsek1995}, which is a simpler 2D model of particles, but capable of describing behaviors ranging from the swarming of small groups moving in random directions, to a global directed motion of the whole population. Recent approaches using more complicated but realistic dynamics for swarming with parameter tuning, include metric distance models with informed leaders \cite{Couzin2005},  zonal interaction models \cite{Lukeman2010}, molecular physics models with geometric and topological interactions \cite{Eriksson2010}, and predator-prey models using radial force laws \cite{Zhdankin2010}.

Most of the typical approaches of modeling collective behavior using dynamical systems, have involved developing models using physical laws or well known mathematical functions that are known to resemble the phenomena in question. Later it is tested or tuned with observations or data, in order to verify whether it resembles the behavior. The opposite approach is to use time series data from experiments to build a full model that fits it as well as possible, using the power of computers. This methodology, commonly referred as system identification, is used in areas like control engineering and econometrics, although most of the available literature is linked to linear models. Some new generic techniques use data to identify nonlinear ODE models \cite{Bongard2007}, while others have focused on inferring the natural laws of physical systems \cite{Schmidt2009}. Efficient computer algorithms for structural ODE building have been proposed especially in biological systems literature \cite{Gennemark2007, Vilela2009}, due to the necessity of scaling the automated building of models to large complex systems involving many variables. The usage of general and flexible modeling frameworks to capture a wider range of complex behaviors from different fields, can also be considered when designing approaches for automated model building. The automated construction of discrete-time models using radial basis functions, have shown the ability to adequately model chaotic dynamics in systems such as infant respiration, strings, and lasers \cite{Judd1995, Judd1996,Small1998,Small2002}.

Due to the complex dynamics observed in collective systems, we suggest taking this data-driven approach. We use computational methods to process time series data and automatically build nonlinear dynamical models that are able to carry out simulations for analysis and prediction. As a first step, we build models from simulated data of the well studied Vicsek model \cite{Vicsek1995}, in order to confirm that our approach is adequate for modeling collective behavior. Real 3D positional data of pigeon flocks \cite{Nagy2010} are then used to construct realistic models capable of performing simulations that emulate the collective dynamics of the data. By evaluating simulations of the retrieved models, new data can be generated and used to perform analysis of the system and quantify hypothesized collective behavioral properties such as the separation, attraction, and speed of the flock.

\section*{Materials and Methods}

\subsection*{Input data}

Both simulated and real experimental data were used as input to build flocking models. The former were generated using the well known Vicsek model \cite{Vicsek1995}, capable of performing 2D simulations of swarming using simple interaction rules. The latter data set was obtained from pigeon flights using GPS devices attached to the pigeons, and has been previously presented and analyzed \cite{Nagy2010}.

\subsubsection*{The Vicsek model}

The Vicsek model \cite{Vicsek1995}, is a simple nonlinear model capable of simulating swarming behavior. The model is essentially a discrete-time system of several particles in a square domain, with their 2D positions updated according to:

\begin{equation}
\textbf{x}_{i} (t+1) = \textbf{x}_{i} (t) + \textbf{v}_{i}(t)
\end{equation}

The velocities have a constant speed \emph{v} and an orientation defined by an angle $\theta$:

\begin{equation}
\textbf{v}_{i} (t+1) = v
\left( \begin{array}{ccc}
\cos( \theta_i(t+1) ) \\
\sin( \theta_i(t+1)  )\end{array} \right)
\end{equation}

The angles of the particles are updated by averaging the trigonometric components of their nearest neighbors:

\begin{equation}
\theta_i(t+1) = \arctan \frac{ \langle \sin(\theta_i(t)) \rangle_r }{  \langle \cos(\theta_i(t)) \rangle_r  }  + \Delta \theta
\end{equation}

In equation (3), $ \langle.\rangle_r$ denotes average of all particles within a fixed radius \emph{r} of particle \emph{i} (including itself) and $\Delta \theta$ is a uniformly distributed random number, which is basically the noise of the system. 
For our study we will consider a slight modification to equation (3), and consider a fixed number of interacting neighbors, inspired by a recent study on the topological distance in flocks \cite{Ballerini2008}. With this change, the model considers a fixed number of nearest neighbors regardless of the separation distance:

\begin{equation}
\theta_i(t+1) = \arctan \frac{ \langle \sin(\theta_i(t)) \rangle_{M,i}}{  \langle \cos(\theta_i(t)) \rangle_{M,i}}  + \Delta \theta
\end{equation}

The modification consists of changing the interaction average to $\langle . \rangle_{M,i}$, which denotes average over the \emph{M} nearest neighbors and itself. Other important parameters are \emph{N} (number of particles), \emph{L} (the linear size of the cell with the particle), and $\eta$ (the range of the noise). The periodic boundary conditions of the original model in \cite{Vicsek1995} were removed for this study in order to have more realistic data, and thus the boundaries of the square cell were extended to infinity to allow continuous motion. From the Vicsek model, we are interested in generating data for cases of high and low density initial conditions (with low noise) since these correspond to a global directed motion of the whole population and the formation of small groups moving in random directions respectively. These two behaviors are good examples of dynamics that can be observed in real collective systems.

\subsubsection*{Pigeon flight data from GPS}

Relevant research in flocking has provided 3D positional data of pigeon flights, with a very fast sampling rate  \cite{Nagy2010}. This data was obtained from lightweight GPS devices attached to ten homing pigeons. The datasets include eleven free flights of pigeons near their roost and four homing flights which basically involve the flock moving from one position to another. In this study we shall consider the data from the four homing flights, due to the simpler flight patterns that are followed.

The data from the four homing flights was further sampled to provide smaller datasets that are easier to handle for a computer algorithm, but still with a rate that is fast enough to capture adequate flight dynamics. With this in mind, sampling rates of one and two seconds were considered depending on the particular properties of each of the flights. In addition to this, the flights were cut to remove idle moments with no significant movement of the birds. Stranded pigeons were also removed from the input data in order to have datasets that resemble a fully interactive population as well as possible. All these edits were made based on thorough manual visualization of the flight data. Some specific details of each of the flights will be explained, since their particular properties will be important to interpret some of the results later on:

\begin{enumerate}
\item
Homing flight 1 (hf1): A flight of 5 pigeons with separations of around 300-350 m from mean position of flock. Initial conditions have some pigeons moving in opposite directions.  Sampling rate: 1 sec.

\item
Homing flight 2 (hf2): A flight of 9 pigeons with separations of around 650-700 m from mean position of flock.  Two pigeons separate from the flock and move together by the end of the flight. Sampling rate: 2 sec.

\item
Homing flight 3 (hf3): A flight of 6 pigeons with separations of more than 1 km from mean position of flock. Sampling rate: 2 sec.

\item
Homing flight 4 (hf4): A flight of 8 pigeons with separations of around 45-55 m from mean position of flock. Sampling rate: 1 sec.

\end{enumerate}

The sampling rates were selected according to the interaction distance between pigeons. That is, for the flights with higher separations (hf2 and hf3), a slower sampling rate of 2 seconds was used, while the flights with closer interactions (hf1 and hf4) were sampled every second. This was done to have a more precise account of the movement when shorter separations are involved in the flight patterns. The different sampling rates also imply that the built models will have to be for a single flight, and this is actually expected, since each flight is different and might contain different terrain and behavioral properties.

\subsection*{Modeling schemes}

After defining the input time series data to use, it is important to select an adequate modeling scheme capable of reproducing the flight dynamics. Recent investigations have led to conclusions about hierarchical structures present in pigeon flights \cite{Nagy2010}, which points to some 'leader' birds having a stronger influence in the decisions of the flock. If we take this into account in our modeling scheme, we would need to build a separate model for each pigeon, and this is certainly possible. Nevertheless, for simplicity we decided to build a single general model for all birds, akin to classical generic models \cite{Vicsek1995,Reynolds1987}, in order to capture and emphasize the general swarming behavior rules that every bird follows.   

To capture a wide range of complex behaviors, black box discrete-time nonlinear dynamical systems can be used to build arbitrary mathematical functions. The obvious disadvantage is that the complex function structure makes it difficult to understand something about the system from simple visualization. Nevertheless, good function fitting properties could make such models very useful for performing simulations and getting conclusions and statistics from this newly generated data. Also of importance is the selection of an adequate embedding scheme for the data to model, which consists in selecting the past values from the data that will be used by the model to predict new positions. The values to consider should be inspired by known physical properties of the phenomena in question (in this case pigeon flights). First the general framework and method used to build the functions of the models will be introduced, and after that, the three different kinds of models to be considered will be outlined.

\subsubsection*{Radial basis functions}

Any competent nonlinear modeling algorithm such as \cite{Bongard2007, Gennemark2007} could be used to fit a dynamical system to the data. We selected the discrete-time radial basis approach originally presented in \cite{Judd1995,Small1998,Small2002} as the modeling framework and algorithm to use, due to its proven capability of modeling highly nonlinear systems. In summary, the method receives as an input a scalar time series scalar time series $y(t)$ and attempts to build the best model of the form:
\begin{equation}
y(t+1)=f(\textbf{z}(t))+ \epsilon(t)
\end{equation}
where $\textbf{z}(t) = [y(t),y(t-1),...,y(t-d)]$ is the embedding of the system and $\epsilon(t)$ is the model prediction error. The former corresponds to the past values from the time series data $y(t)$ that the model will consider for calculating the prediction for $t+1$. The samples used for optimization of the model are built from time series $y(t)$ using the embedding $\textbf{z}(t)$. The structure of the function to build follows:
\begin{equation}
f(\textbf{z}(t)) = \lambda_0 + \sum_{i=1}^{n} \lambda_{i} y(t-l_i) + \sum_{j=1}^{m} \lambda_{j+n} \phi_j \left( \frac{\| \textbf{z}(t) - c_j \|}{r_j} \right)
\end{equation}
with $r_j$ and $\lambda_j$ as scalar constants. The $c_j$ parameters denote random points, known as centers of the radial basis functions. The first sum (and constant $\lambda_0$) in (6) is the linear part of the system, equivalent to commonly used autoregressive models (AR). The second sum is the nonlinear part of the function, and it is characterized by radial basis functions $\phi_j$, which can be of different types, as shown in \cite{Small1998}. The algorithm that builds and optimizes the model, described in detail in  \cite{Judd1995,Small1998}, basically consists in generating random candidate radial basis functions, choosing the ones which best follow the data one by one, and estimating the parameters.  The optimization for selecting the best model is done by minimizing the Minimum Description Length (MDL) \cite{Rissanen1989} to prevent overfitting of the data.

\subsubsection*{Relative position modeling for Vicsek data}

The first and simplest model to be built is the one based on the data of the Vicsek model. These models will be referred as type \textbf{R1}, symbolizing the first variant of \emph{relative} models. As mentioned in the introduction to this section, the idea is to use the same general model for each particle of the system. Since the Vicsek data is two-dimensional, we require two different functions for a single model (to predict each coordinate). Therefore the single model that all individuals follow can be defined as:

\begin{equation}
\textbf{f}[\textbf{z}(t)]=
\left( \begin{array}{ccc}
f_1(\textbf{z}(t)) \\
f_2(\textbf{z}(t))\end{array} \right)
\end{equation}

Since the Vicsek model considers relative positioning (the numerical positional value of a particle does not influence its movement), instead of trying to predict the absolute position $\textbf{x}_i$ at $t+1$, we can re-define equation (5) by predicting the relative change in position $\Delta \textbf{x}_i (t+1) = \textbf{x}_i(t+1)-\textbf{x}_i(t)$.
\begin{equation}
\Delta \textbf{x}_{i}(t+1) =\textbf{f}[\textbf{z}_i(t)]  +  \textbf{e}_i(t)
\end{equation}
where $\textbf{e}(t)$ is an array with the model prediction errors for each coordinate. To capture the collective behavior of the source system in our model, the embedding $\textbf{z}(t)$ of a particle \emph{i} should consider enough information from the nearest neighbors that influence its motion. An adequate embedding would be to consider the average change in position $\Delta \textbf{x}_i(t)$ of \emph{i} and its neighbors, since it gives enough information about the magnitude and direction of velocity $\textbf{v}_i(t)$, which actually is enough to model the Vicsek rules (excluding the noise, see equations 1-4). At a contrast to the original temporal embedding of the radial basis method, this new embedding considers data from the nearest neighbors, but it does not affect the modeling algorithm since the method fits a function to emulate the output samples from given embedding instances (inputs), regardless of the embedding form. Taking this into consideration, the embedding would be different for each particle due to the difference in nearest neighbors:
\begin{equation}
\textbf{z}_i(t) =\begin{array}{ccc}
\langle \Delta \textbf{x}_i (t) \rangle_{M,i}  \end{array}   
\end{equation}
Of importance here is that the neighborhood of the average in (9) considers the fixed-number interaction introduced previously for the modified Vicsek model in (4), which is the averaging over particle \emph{i} and a fixed number of \emph{M} nearest neighbors. Also of relevance is that positions are two-dimensional, and thus the embedding in (9) consists of two variables.

\subsubsection*{Absolute position modeling for pigeon homing flights}

When modeling the real 3D homing flights, we must take into account that the pigeons are following a trajectory, which can be summarized as a flight from point \emph{A} to point \emph{B}, with some terrain information on the way that will influence their flight patterns. This means that for adequate modeling of a homing flight, we require absolute positioning in our model. In other words, at a contrast with the Vicsek model, the position $\textbf{x}_i(t)$ of pigeon \emph{i} is necessary for an adequate prediction of its value at $t+1$, due to the terrain information being absolute. Equation (10) shows the absolute model structure to be used.
\begin{equation}
\textbf{x}_{i}(t+1) =\textbf{f}[\textbf{z}_i(t)]  +  \textbf{e}_i(t)
\end{equation}
This model type will be referred as \textbf{A}, symbolizing the word \emph{absolute}. It is also worth mentioning that since positions are now three-dimensional, the model consists of three functions instead of the two that were required for the Vicsek data (7). The usage of experimental data requires a more complete embedding for adequate modeling. First of all, physical common sense should be taken into consideration to define a model. That is, at least second-order components must be considered, which translates into the necessity of using velocity information (values at time $t-1$ and $t$) to calculate a prediction for $t+1$. In addition to this, we are interested in modeling the collective behavior, and thus neighbor interactions must be included. Taking all of this into consideration, the proposed embedding is displayed in (11).
\begin{equation}
\textbf{z}_i(t) = \left( \begin{array}{ccc}
\textbf{x}_i(t) \\
\textbf{x}_i(t-1) \\
\langle \textbf{x}_i(t) \rangle_M \\
\langle \textbf{x}_i(t-1) \rangle_M  \end{array} \right)    
\end{equation}
This embedding scheme for a pigeon \emph{i} considers its absolute position at the two previous time intervals ($t$ and $t-1$) and the average position of its \emph{M} nearest neighbors in the same intervals. Note that each of the four components in (11) is three-dimensional, which translates into a 12 variable embedding, making it much more complex than the two variable embedding used for the Vicsek data. Another distinction from the Vicsek embedding, is that the nearest neighbor averaging in (11) considers only the \emph{M} nearest neighbors and not itself. This is a reasonable thing to do, since the position of pigeon \emph{i} is already being directly considered in the embedding, taking into account the more realistic consideration of a significant difference between its own position with that of its neighbors. From this, we will denote  $[\textbf{x}_i(t), \textbf{x}_i(t-1)]$  and $[\langle \textbf{x}_i(t) \rangle_M, \langle \textbf{x}_i(t-1) \rangle_M]$ as the \emph{individual} and \emph{collective} components of the model respectively.
  
\subsubsection*{Relative position modeling for general flocking model} 

The modeling scheme presented in the previous subsection was designed for navigational flights, where absolute positioning is important due to the influence of the terrain in the flight. In addition to these models, it is of our interest to build a general flocking model using relative positioning similar to the Vicsek model, but based on real experimental data. This final model type shall be referred as \textbf{R2}, which refers to the second variant of \emph{relative} models. As a first step, we should mention that the desired relative position model structure is the same as in (8), and introduce a new nine-variable embedding:
\begin{equation}
\textbf{z}_i(t) = \left( \begin{array}{ccc}
\Delta \textbf{x}_i(t) \\
\langle \Delta \textbf{x}_i(t) \rangle_M \\
\textbf{x}_i(t) - \langle \textbf{x}_i(t) \rangle_M \end{array} \right)    
\end{equation}  
Here the first component represents the positional change of bird \emph{i} at time \emph{t}: $\Delta \textbf{x}_i (t) = \textbf{x}_i(t)-\textbf{x}_i(t-1)$, the second component is the average positional change of the nearest neighbors, and the third component is the averaged positional difference between \emph{i} and its nearest neighbors. The first two components resemble the Vicsek embedding introduced in (9), though now separating the change of bird \emph{i} from that of its neighbors. The third component symbolizes a directional separation between \emph{i} and its neighbors, which should be useful to characterize collective behavior. For example, we can expect that the separation distance to its neighbors in some direction (front, back, left, right) will surely have an effect on the movement of bird \emph{i}. This component introduces a dependence on the metric distance between neighbors, which makes the model a hybrid inspired by both topological and metric distance approaches (see \cite{Ballerini2008}). Essentially the metric separation distance to its \emph{M} nearest neighbors will influence an individual's movement. 

We must emphasize that the plain homing flight data is not appropriate for building this general flocking model, due to the navigational bias that it has. To exemplify this, the four homing flights previously introduced consist of pigeon flocks moving from a point \emph{A} to \emph{B} in a loosely southwest direction. This means that if data of one or all the flights is used to build the model, it will undoubtedly be biased with southwest movement. In order to eliminate the bias as best as possible, we performed uniform 2D rotation transformations to produce new embedded data $\textbf{z}_i (t)$ and prediction values $\Delta \textbf{x}_i (t+1)$. For simplicity, these rotations were only done for the latitude and longitude coordinates, leaving the altitude component intact. Figure 1 shows a graphical example of a -90 degree rotation of a single embedding and prediction instance. Each rotation angle for an instance of (12) is calculated so that the orientations of the prediction, $\Delta \textbf{x}_i(t+1)$, span a full circle ($2\pi$) in the whole dataset. These transformations are done with respect to $\Delta \textbf{x}_i(t+1)$ because this vector is the actual navigational force of the model, which has the previously mentioned bias in the original data. In summary, by spanning a full circle in the navigational direction of the samples, we are attempting to attenuate the bias.  


\subsection*{Measuring flock dynamics}

As a valuable tool for the analysis of flocking data, a measure that can characterize and illustrate the dynamics of the collective behavior should be used. Velocity correlation functions have been considered in previous studies \cite{Okubo2001} to calculate the positional variation of individuals in swarming populations. We propose a generic measure to illustrate the dynamic behavior of a flock with a single time-dependent variable, by defining the average separation of individuals from the mean position of the whole population (\emph{N}) at a time interval \emph{t} as:

\begin{equation}
\delta_g(t)=\frac{1}{N} \sum_{i=1}^{N} \| \textbf{x}_i(t)-\langle \textbf{x}(t) \rangle_N \|
\end{equation}

This measure should be relevant when analyzing the global dynamical properties of a whole flock, but it could also be important to measure the separations in local neighborhoods instead of the whole population, especially when swarming in small groups is significant. A slight modification to (13) gives us the average separation of individuals from the centroid of their local neighborhood of neighbors:

\begin{equation}
\delta_l(t)=\frac{1}{N} \sum_{i=1}^{N} \| \textbf{x}_i(t)-\langle \textbf{x}_i(t) \rangle_{M,i} \|
\end{equation}   

Besides the analysis of each data set, the main idea of introducing these measures is to perform qualitative comparisons between the $\delta$'s of the input data and data obtained from model simulations, in order to verify the dynamics of a retrieved model. These comparisons could be more relevant than a quantitative verification, e.g. the shape of the curves having more importance than the absolute error between them, for emulation of collective behavior. Another important feature of using these time-dependent measures is that transient and steady state properties of the system can be easily visualized, in a way that is very similar to analyses commonly done in control theory. Essentially a stable steady state in the $\delta (t)$ signal represents ordered synchronized movement of the whole flock (or a small group if considering the local measure). For some cases, it might be more convenient and compact to express the separation of a system as a single numeric quantity, and therefore the average separation over all time intervals \emph{T} will be defined as $\overline{\delta}_g=\langle \delta_g(t) \rangle_T$.

\subsection*{Methodology}

The automated modeling process shall now be outlined. As described in section 3, radial basis functions are used to build three different model types (R1, A, R2), according to the introduced modeling schemes.  Essentially the same procedure is followed to build the three types of models, with the notable differences being the input data, number of functions per model, and embedding. The general process to build a single model will be described, with additional special comments for each model type:

\begin{quote}
Input: Time series matrix containing positional data: $\textbf{x}(t)$ 
\begin{itemize}
\item Vicsek model simulations for R1 models
\item Homing pigeon data from a single flight for A models
\item Homing pigeon data from multiple flights for R2 models
\end{itemize}

For each positional coordinate \emph{j}:
\begin{enumerate}
\item
Build samples for the function $f_j$ using as output $\Delta x_{i}^{(j)} (t+1)$ (for R1 and R2 types) or $x_{i}^{(j)} (t+1)$  for (A types) with their respective embedding $\textbf{z}_{i}(t)$, using data from all individuals/pigeons.
\item
Run the radial basis modeling algorithm.
\item
Set the retrieved function as $f_j$.
\end{enumerate}
\end{quote}

The randomness in the radial basis modeling algorithm used to construct the models (see \cite{Judd1995,Small1998}), makes it necessary to run the algorithm several times to retrieve several models and find the most appropriate one, or average their statistics. Specific details on the number of retrieved models, and model selection criteria will be discussed in the results section.

\section*{Results}

Using the modeling schemes presented in section 3 and the procedure from section 5, several models were retrieved for each dataset. To analyze the models, the main philosophy followed in this study was to perform simulations with different sets of initial conditions and verify their behavior using the measures introduced in section 4, by either comparing with the source input data or simply by analyzing the simulated data itself. More specific details for each model type shall be discussed in each subsection.


\subsection*{R1 models (Vicsek model data)}

From samples of five different instances of low density simulations of the modified Vicsek model ($L=25, \eta=0.1$ see \cite{Vicsek1995}, $M=4$) , five different R1 models were obtained following the procedure from section 5. From there, we decided to select a single ``best" model by comparing the global behavior under high density conditions ($L=5, \eta=0.1$ \cite{Vicsek1995}, $M=30$), which was done by averaging $\overline{\delta}_g=\langle \delta_g(t) \rangle_{K,T} $ over ten simulations (\emph{K})  of high density initial conditions, each with 500 time intervals (\emph{T}). The model with least absolute error: $|\overline{\delta}_g^{(data)}-\overline{ \delta}_g^{(model)} |$ was selected. With this criteria, we chose to discriminate the models by comparing their extrapolating capabilities to simulate initial conditions with different density than their input data, i.e. the phase transition in the classic Vicsek model. The number of nearest neighbors (\emph{M}) in the model structures and for calculating  $\delta_l(t)$ were chosen to be $M = 4$ for low density initial conditions and $M = 30$ for the high density case, which from simulations resemble the steady state number of neighbors from the original radius interaction in the Vicsek paper \cite{Vicsek1995} \footnote{The original Vicsek model with radius nearest neighbor interaction was also used as input data for a separate test of R1 model retrieval. The results were very similar (only slightly worse) due to the loosely constant number of steady state neighbors in the original model.}.

In figure 2 we can see a comparison between the global separation dynamics of the modified Vicsek model and the R1 model. The retrieved model closely emulates the behavior of the source model. The low density cases show an expected higher separation rate than the high density simulations, which is related to swarming occuring in small groups and moving away. Even for the extrapolating case of high density initial conditions, the model exhibits close following of the dynamics.  
Figure 3 shows a comparison of the local neighborhood dynamics of the same two models. In (a), The Vicsek data reaches a pseudo-steady state for low density initial conditions, with the R1 model closely following it but reaching a more stable state. Nevertheless, the deviation is minimum, and likely caused by the noise in the source data affecting the long range interactions. The high density case in figure 3(b) considers closer interactions, and thus both models reach a stable state, with a small transient in the source model, and a small steady state error.   


As inferred with the $\delta_g (t)$ curves, the low density initial conditions of the Vicsek model feature swarming in separate groups moving in different directions. Figure 4 shows snapshots of simulations of both models for the same low density initial conditions. Qualitatively the retrieved model follows the behavior of the source model quite well, and this was verified through several simulations. The only noticeable issue observed through the simulations is a slightly biased default direction followed by some individuals, but this should be expected when constructing a model with imperfect data. Using more data to build the model, this bias is removed, but the trade-off is longer computational time. Nevertheless, the observed bias was small, and many different directions were still seen in the simulations. The same conclusions were observed for the high density simulation comparisons, which involves ordered movement of the swarm in only one or two big groups. Figure 5 shows how a split in the population was emulated well qualitatively by the R1 model, but with an alignment difference. This modeling scenario of the Vicsek data and the results, are a good introduction to the task of building models from real experimental data, which is noisy and even more imperfect, as shall be considered in the next subsections.


\subsection*{A models (Homing pigeon flight data)}

The main purpose of the A models is to emulate their respective flight (input data) through simulations, and to confirm that there is collective behavior influencing the models and not simply an individual navigational force. In order to see the effect of the collective components and the interaction structures, models with different number of nearest neighbors (\emph{M}) were obtained for comparison. In addition to those, models with no collective components ($M=0$) and thus a six variable embedding (see 11), were also retrieved for the same purpose. To obtain better conclusions of the collective effects, five different 3D models were obtained for each value of \emph{M} considered, and all statistics averaged over the five. For the model analysis, two different simulation scenarios were considered for each homing flight: same initial conditions as the input data and random initial conditions.  The latter were calculated from a normal distribution with the same mean and variance as the absolute initial positions of the input data (for each of the three coordinates), in order to preserve similar flight conditions, but with no initial velocity, ($\textbf{x}_i(1)=\textbf{x}_i(0)$). The global separation measure $\delta_g (t)$ calculated in the simulations, was used as the comparison statistic between model structures and input data.   


For simulations of the homing flight 1 (hf1) models (from $M=0$ to $M=4$) with five pigeons, figure 2(a) shows that models with low \emph{M} (0, 1 and 2) in average do not follow the input data for the same initial conditions. This deviation in $\delta_g(t)$ was observed to happen early in the simulations at $t < 100 s$ and the opposite velocities in the initial conditions of some of the individuals (as introduced in section 2.2) is what likely causes this difficulty. Nevertheless, the models with higher \emph{M} (3 and 4) follow the input data neatly (figure 2(a) shows $M=3$). When considering random initial conditions with no initial velocity, in figure 2(b) we can see how the individual models ($M=0$) do not keep cohesion of the flock as good as the collective models. This confirms that a collective force is modeled, and that an interaction structure with $M=3$ offers both accurate path simulation and more cohesion \footnote{The M=4 curves are not displayed in figure 6 for better visualization; they follow M=3 closely.}. 


The homing flight 2 models present probably the most difficult case to analyze. As can be seen in figure 7(a), the input data considers the separation increasing at around $t > 500$, and this happens because the data has two pigeons moving together away from the main flock at that time. This causes an interesting modeling case, since these particular data contradict the cohesive tendencies found at $t < 500$. All this has repercussions with the retrieved models. Figure 7(a) shows how with the input data initial conditions, the individual models ($M=0$) follow this separation better than the others. This can be reasoned with the fact that it is purely using positional information to estimate the trajectory, and the small variations between positions at $t \approx 500$ are what cause the divergence that mimics the input data. Nevertheless, the collective models are again expressing a significant collective force, since they give preference to the group tendencies and thus do not follow the separation increase. To better visualize how the interaction structure affects the collective component for this flight, $\delta_g(t)$ was averaged over all time intervals for each model structure. Figure 7(b) shows that the simulations with random initial conditions found the most cohesive interaction structure to be at $M=4$, and a surprising separation for $M>4$. This likely stems from the fact that the two-bird deviation in the input data causes models with larger interaction neighborhoods to be more sensitive to the deviations of some individuals. Overall, this particular flight illustrates how the input data can also affect what the models will try to capture, as well as the trade-offs between modeling the navigational trajectory or the collective behavior more closely.        


At a contrast to the previous two homing flights, the third one has separations of up to 1 km between pigeons, and thus it considers the longest interaction range of all the flights. The number of pigeons in this data set is 6, and models with $M=0$ to $M=4$ were retrieved. It was found that every single model structure follows the input data closely when using the same initial conditions. Figure 8(a) shows it for $M=0,1,3$. When using random initial conditions, no discernible pattern was found for the average global separation $\overline{\delta}_g$ as a function of nearest neighbors, with all values very close to 1.1 km. This confirms that the large separations in this flight make the navigational components be the only driving force of the individuals. This means there is no significant collective behavior for this flight, and an individual model with $M=0$ should be as good as any to simulate the dynamics.      


Finally we arrive with homing flight 4, which considers eight pigeons with the shortest average separations of them all, at around 50 m. This causes $\delta_g(t)$ to appear quite noisy due to the expected sensitivity of separation distances in a higher density flock. With \emph{M} ranging from 0 to 6, we found that the models with highest value at $M=6$ produced both the best following of $\delta_g(t)$ for the input data initial conditions and the highest cohesion when simulating random initial conditions. Figure 5 shows a comparison of models with $M=0,3,6$. At a difference with the previous three flights, for the random initial conditions we considered a normal distribution with two times the standard deviation of the input initial conditions, instead of one. This was done in order to start with larger separations and verify the dynamic attraction properties of the models. From the results, all the collective models clearly had more attraction, and in figure 5(b) we can observe how two of them show sharp convergence tendencies that the individual models do not have. 

In general, from analyzing the results of the four homing flight models, we can confirm that our approach is adequate for predicting and simulating pigeon flock trajectories. For flights with close interaction ranges (less than 700 m from mean position per bird), the collective models can better capture the flight properties and offer the best flock cohesion when changing initial conditions. For long interaction ranges (around 1 km), simple individual models that consider only own positional information, were enough for a simulation of the flight. 

\subsection*{R2 models (Homing pigeon flight data)}

The data from hf1, hf2, and hf4 was used with the modeling scheme outlined in section 3.4 to build five models for each structure with $M=1$ to $M=4$, and using two second sampling on the data. The third homing flight was excluded since our results in section 6.2 confirm there is no strong collective behavior in that dataset, and \emph{M} was limited to a maximum value of 4 because of the hf1 data that only considers five pigeons. At a contrast with the A models, the R2 models were set to be 2D for simplification, with the height component removed from predictions and simulations. To select the ``best" model, ten different sets of initial conditions that resemble the properties of hf1 (mid-range interaction around 300 m) were used to simulate flights of $N=9$ individuals, and $\overline{\delta}_g$ measures calculated. The flight with highest average cohesion (lowest $\overline{\delta}_g$), was selected as the ``best model" due to its higher flocking capabilities; and not surprisingly it was a model with $M=4$, since $\overline{\delta}_g$ was found to decrease as \emph{M} increases. This model was used for all the analysis and it shall now be referred as the R2 model in general terms. Also important to note is the fact that the R2 model has a slightly biased overall direction, which was reduced as best as possible using the rotational modeling scheme presented in section 3.4. 


Many different variations of initial conditions or even values of \emph{M} and \emph{N} could be considered on the R2 model simulations for analysis. For this study we decided to fix \emph{N} and \emph{M} at 300 and 4 respectively, and vary the initial positional density of the individuals (just like in section 6.1), and the initial speeds (the magnitudes of $\Delta \textbf{x}_i (1)$). The directions of $\Delta \textbf{x}_i (1)$ were obtained from a random uniform distribution $U(0,2\pi)$. The initial density was varied according to the formula $r=250r_c$, where \emph{r} is the radius (in meters) of the circle in which the initial positions of the individuals are distributed (following a uniform random distribution), and $r_c$ is the actual coefficient that is varied. The initial speed for each particle was calculated from $v_i=30v_c$, with $v_c=[0,1]$. Considering that the R2 model has two-second updates (input data with two second samples), this limits the initial speed to a maximum of 15 m/s, which is roughly near the average of the speeds in the pigeon homing flights. Seven different $r_c$ values and eleven $v_c$ values were considered for a total of 77 combinations of initial condition parameters. Each parameter setting was considered for ten different simulations, in order to get better averaged statistics.


In figure 10, we can see a comparison of $\delta_g(t)$ for four extreme cases of high and low densities and velocities. Figure 10(a) shows that cases of low velocity settle approximately into steady states, while the high velocity cases have increases in global separation; more drastically in the high density case. This tells us that global flocking is highly dependent on the velocities of the individuals, and not so much in the population density. In figure 10(b), the local separation properties can be observed. Basically for all cases, the individuals tend to converge into local groups with less than 20 m separation, with the low density cases having drastic drops in $\delta_l(t)$ that symbolize strong attraction. This means that larger separations provoke individuals to move strongly toward their neighbors. The high density and high speed case shows an interesting initial increase in $\delta_l(t)$ and then decrease to settle down into its steady state. This is likely due to velocity synchronization: individuals are initially close together with random directions but later on they separate and align with neighbors with similar orientations and synchronize their velocities.


Figure 11 shows how in the low density and low speed simulation, the individuals attract into small groups and globally move together in the main direction (the biased direction: roughly southwest). For the high speed case, the flock has less cohesion, including some stranded individuals moving in other directions, and therefore their alignment is not fully synchronized. From here we can see that when individuals are separated by a longer distance, if they are moving at slower speeds then they have a better chance of finding each other and aligning. The higher speeds make it more difficult, and thus provoke less cohesion and stranded groups of individuals moving on their own. These tendencies are consistent with figure 10. For the two cases of high density initial conditions we have a drastic difference. Figure 12 shows how for low speeds, the flock stays together, spaces out, and then slightly moves in the main direction. For high speeds, small groups are formed and they move away from the center independently. This shows that the velocities are usually roughly maintained within nearest neighbors, and thus cause the significant difference in system behavior. 


To generalize how the initial densities and speeds affect the system behavior, figure 13 shows a plot of the averaged separation $\overline{\delta}$ in shades of gray, for the different initial parameters. In figure 13(a), we can see how by increasing the speed, the separation increases for every density value, though in a lesser rate for the low density cases. For a fixed speed value, when decreasing the density (increasing the radius coefficient $r_c$), the separation tends to decrease and then increase again, which implies there is a critical density value with highest cohesion for each different speed value (marked on the figure). As the speed increases, a lower density (higher $r_c$) value will be required to achieve highest cohesion. Figure 13(b) shows how the local separation follows the expected pattern of higher cohesion at higher densities and lower cohesion at lower densities, with no strong dependency on the speed. Relevant to notice is that the cohesive force does not decrease so drastically until it passes a distance threshold of no interaction between individuals. From these observations, we can conclude that speed has a higher influence on the global dynamics of the system, while density influences the local interactions. In other words, the density seems to have a greater influence on the directional alignment of individuals: lower separations causing almost immediate neighbor alignment, while larger separations require a transient to first converge and then align (see figure 10(b)). 


As final illustrations of the behavior of the R2 model, averaged distributions of how the separation of a particle \emph{i} from its nearest neighbors affects its attraction and speed at the next time interval, were calculated from all of the simulations considered in this subsection (all 77 combinations of parameter values and their ten simulations). The average nearest neighbor separation of a particle \emph{i} at a time interval is defined as $\xi_i(t)=\frac{1}{M} \sum_{j=1}^{M}   \| \textbf{x}_i(t) - \textbf{x}_{nn_j}(t) \|$. Figure 14(a) shows that there is a small repulsion force spanning up to slightly less than 20 m of separation and then the strong attraction force with a maximum strength at around 120 m and ending near 500 m, which tells us that the maximum interaction range is approximately 500 m. The distribution is much smoother for short range interactions, likely because cohesive scenarios dominated the simulations. Figure (b) shows how the speed of a particle is highest within 20 and 120 meters of separation, with a maximum near 35 m. This region is strongly correlated with the strong attraction region also roughly covering that span until the minimum (maximum attraction) is reached in figure 14(a). For higher separations, the speed decreases steadily, which follows that individuals are attracted less to their neighbors as their interaction decreases. The maximum speed at 35 m presents an interesting interpretation, because it only amounts to a weak attraction force, implying an interaction range where individuals are aligned and moving fast together.

\section*{Discussion}

The radial basis modeling algorithm and the methodology presented here are capable of obtaining models of collective systems exhibiting swarming properties, from which simulations and statistics can be obtained. The fitting of radial basis functions is similar to a recent method using Gaussian processes to model pigeon trajectories and identify terrain landmarks \cite{Mann2010}, with the main difference being that we build a multi-agent model that emphasizes the collective dynamics of a flock instead of a model for the trajectory followed by a single pigeon. The Vicsek model served as an introductory example of a simulated system with noisy data, to verify the efficiency of our approach on modeling dynamics with collective behavior. The results showed adequate qualitative emulation of the dynamics for two extreme density scenarios, even when the sampled data used to build the model was only based on a single density case. This capability of extrapolation on a data set with different conditions also motivates the usage of this approach for real experimental data with more sensitivity and noise. When modeling the four homing pigeon flock trajectories, models with different interaction structures were retrieved in order to test and verify both the trajectory and the collective dynamics. The best retrieved models showed the capability of qualitatively simulating the trajectory from the input data, when using the same initial conditions. By using random initial conditions, it was possible to verify the collective components in the models. The best were found to have the capability of balancing between having the best trajectory simulation and flock cohesion, especially in the flights where collective interactions form an integral part of the dynamics. This also confirmed the adequacy of using fixed-number neighbor interactions to characterize pigeon collective behavior, as well as using a single model for all the pigeons.

The final general flocking model (R2), built upon rotated data of the three pigeon flights with significant collective behavior, showed the ability to represent several swarming behaviors. In general, it illustrated that global swarming of a population is largely dependent on the speeds of the individuals, with low speeds favoring unified organized movement, and high speeds causing local swarming and separation in different directions. The population density also had an effect on the simulations, by essentially having a critical value of highest cohesion for a fixed value of initial velocities. This implies that for given velocities, a certain density distribution will offer the best global cohesion and synchronization between individuals. Finally, the simulations of this model were used to average attraction and speed distributions based on the separation of a particle to its nearest neighbors. The former plot was consistent with classical swarming models that consider a short range repulsion force followed by attraction at longer ranges within the sphere of interaction of an individual. This interaction range between the fixed number of nearest neighbors was found to have a maximum attraction force near 120 m of separation and its limit at around 500 m for this general model built upon pigeon data.

From these observations we can conclude that lower speeds facilitate pigeon flocking, and this is consistent with common sense, since the pigeons can easily converge and synchronize with no major effort. Higher speeds will cause only pigeons with similar alignments to synchronize their directions and move together, which is synonymous to flocking in small groups. Nevertheless, certain densities can still cause global flocking with high speeds, as long as there is enough space for these small pigeon groups to converge and synchronize their velocities.

\section*{Acknowledgments}

The authors would like to thank M\'{a}t\'{e} Nagy who kindly provided the trajectory data of the pigeon flocks.
GDK is currently supported by the Hong Kong PHD Fellowship Scheme (HKPFS) from the Research Grants Council (RGC) of Hong Kong. XKX is currently supported by the PolyU Postdoctoral Fellowships Scheme (G-YX4A) and the Research Grants Council of Hong Kong (BQ19H). XKX also acknowledges the Natural Science Foundation of China (61004104,
60802066).

\bibliography{refs}

\section*{Figures}

\begin{figure}[!ht]
  \centering
  \subfloat[Original data example]{\label{fig:embed1}\includegraphics[scale=0.50]{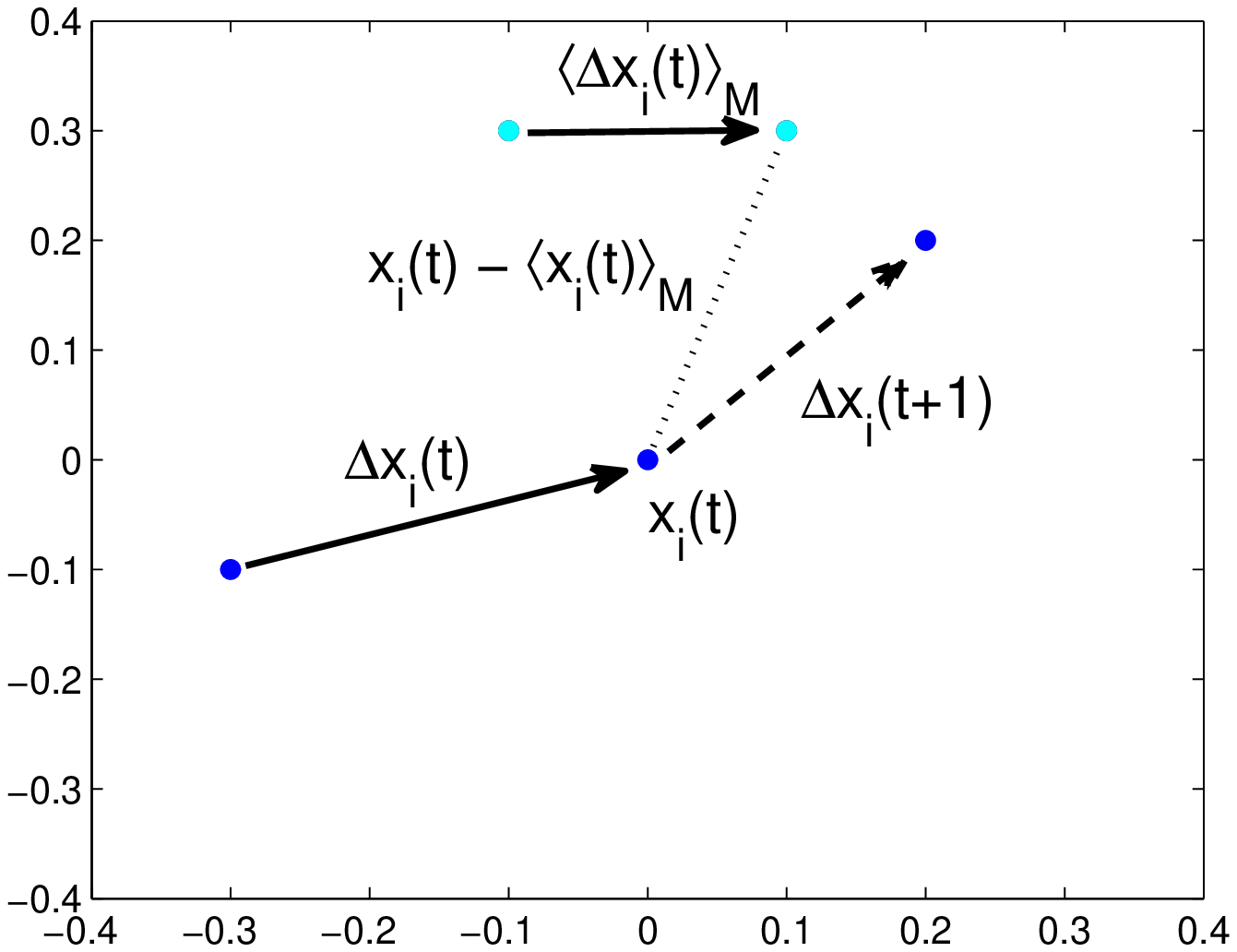}}                
  \subfloat[Rotated data]{\label{fig:embed2}\includegraphics[scale=0.50]{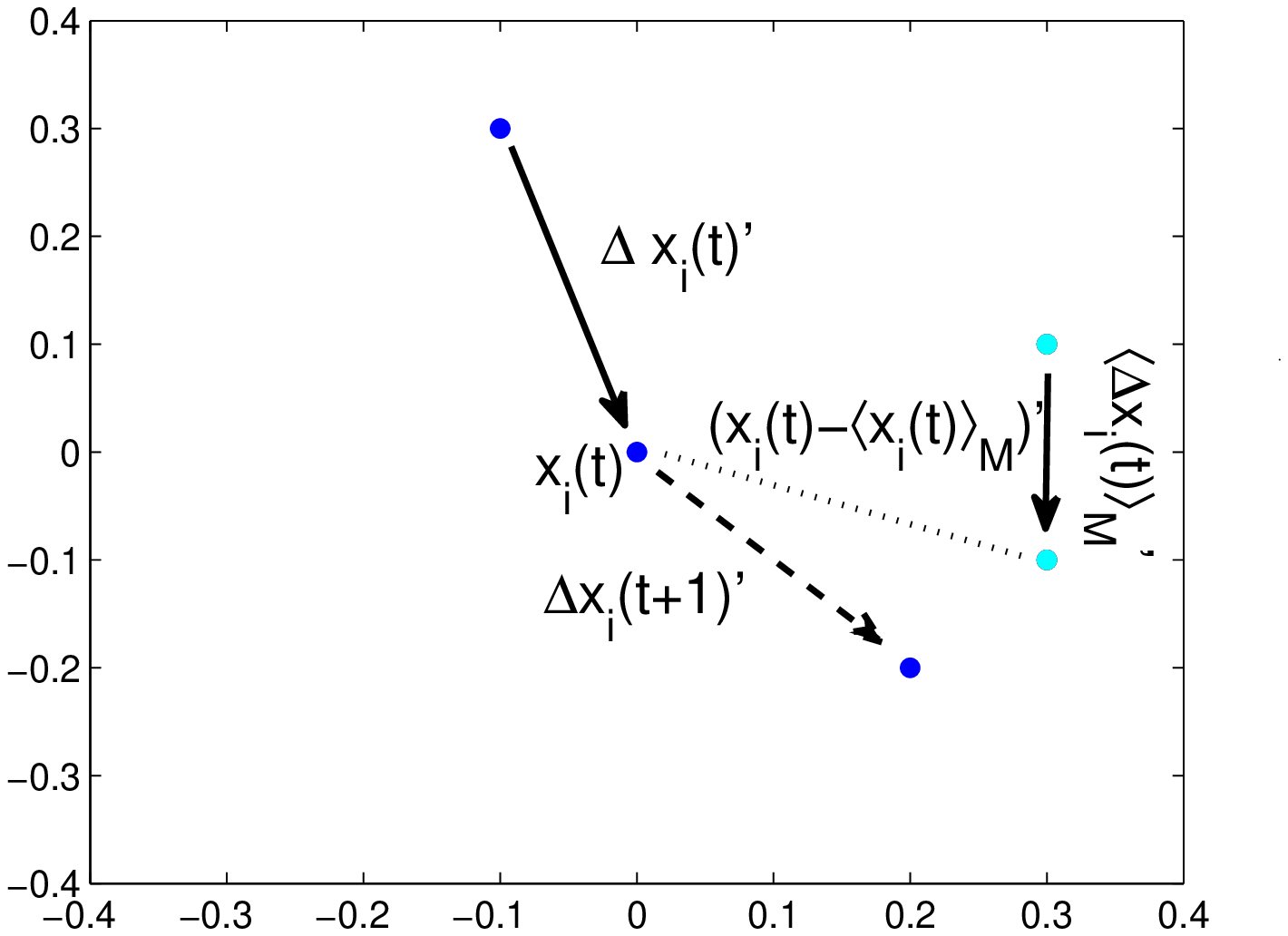} }
  \caption{{\bf A graphical example of a $\mathbf{-\frac{\pi}{2}}$ rotation of embedding and prediction.} In (a) we have the hypothetical original data, and (b) the new data}
  \label{fig:embed}
\end{figure}

\begin{figure}[!ht]
  \centering
  \subfloat[Low density initial conditions]{\label{fig:vm1}\includegraphics[scale=0.50]{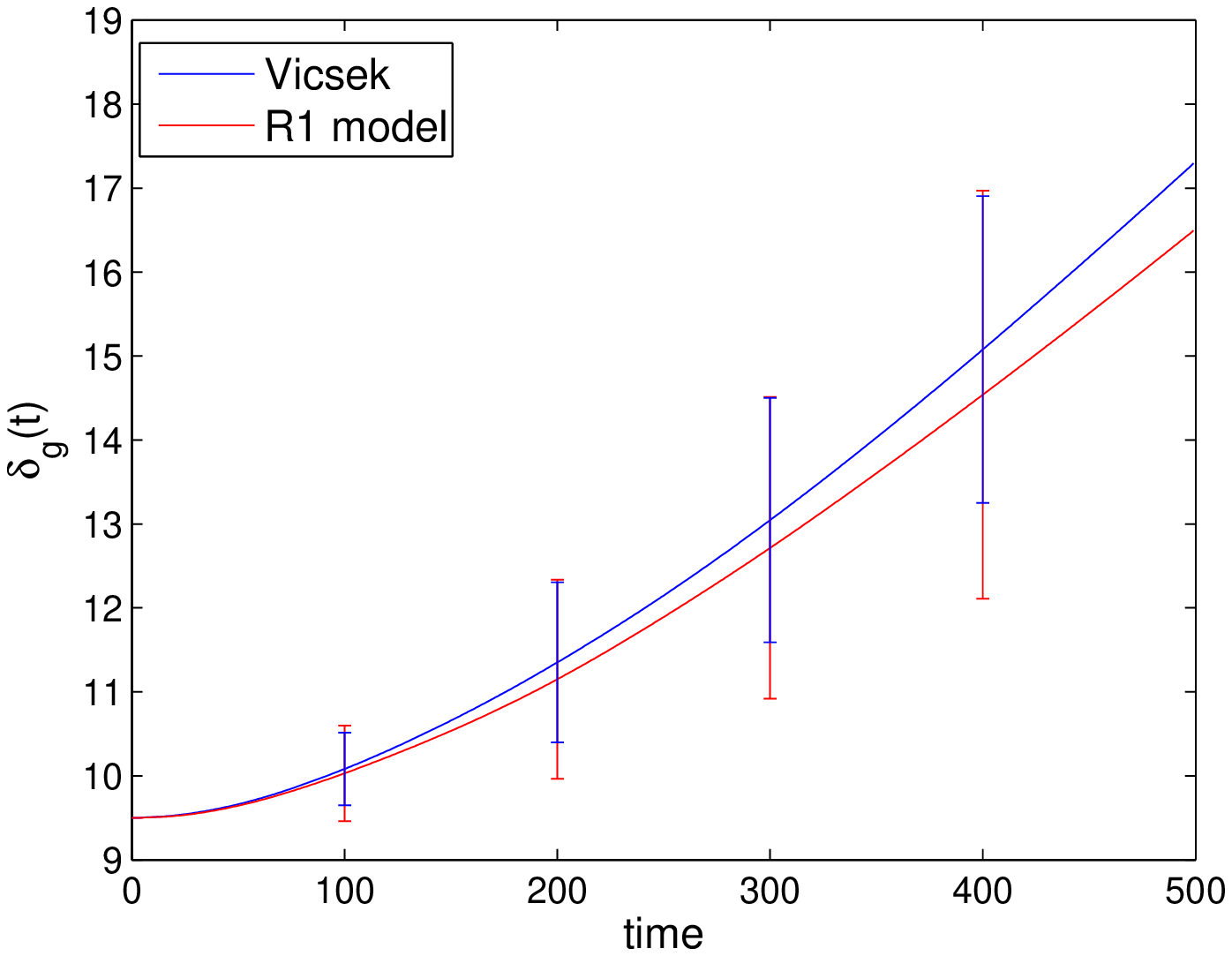}}                
  \subfloat[High density initial conditions]{\label{fig:vm2}\includegraphics[scale=0.50]{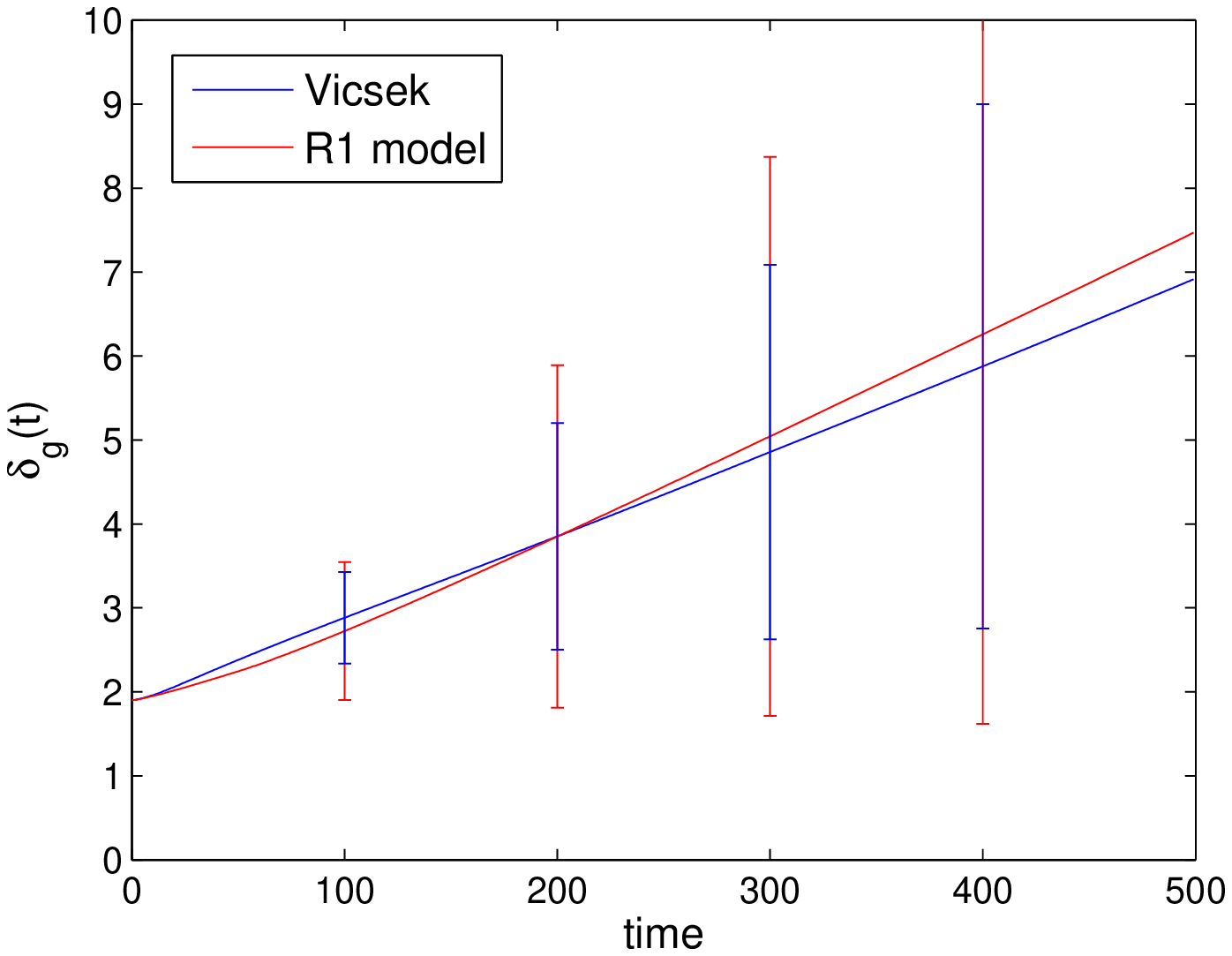} }
  \caption{{\bf Comparison of $\mathbf{\delta_g(t)}$ between the modified Vicsek model data and the ``best" R1 model.} Statistics were averaged over 10 simulations, with both models using the same initial conditions.}
  \label{fig:vmg}
\end{figure}

\begin{figure}[!ht]
  \centering
  \subfloat[Low density initial conditions]{\label{fig:vm3}\includegraphics[scale=0.50]{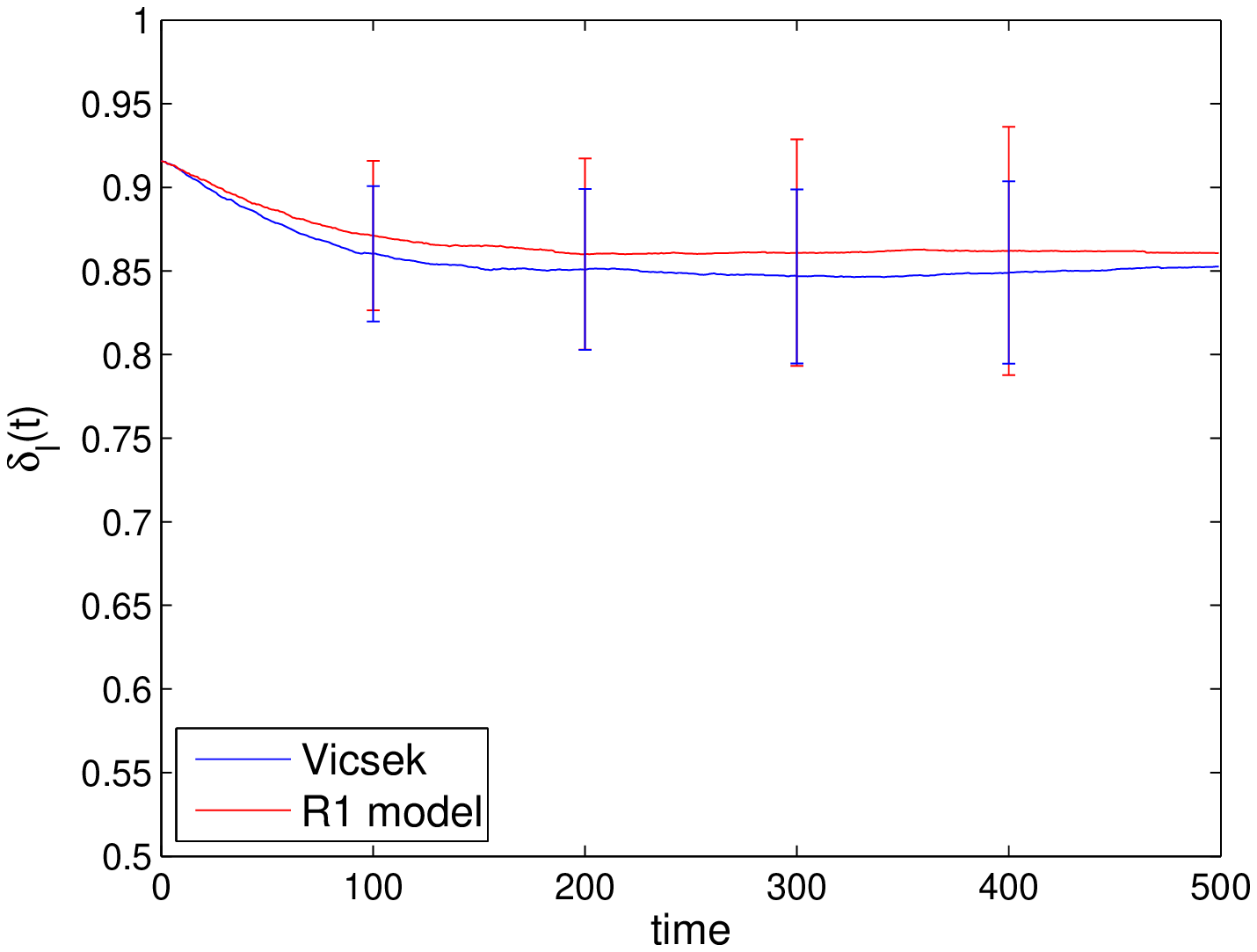}}                
  \subfloat[High density initial conditions]{\label{fig:vm4}\includegraphics[scale=0.50]{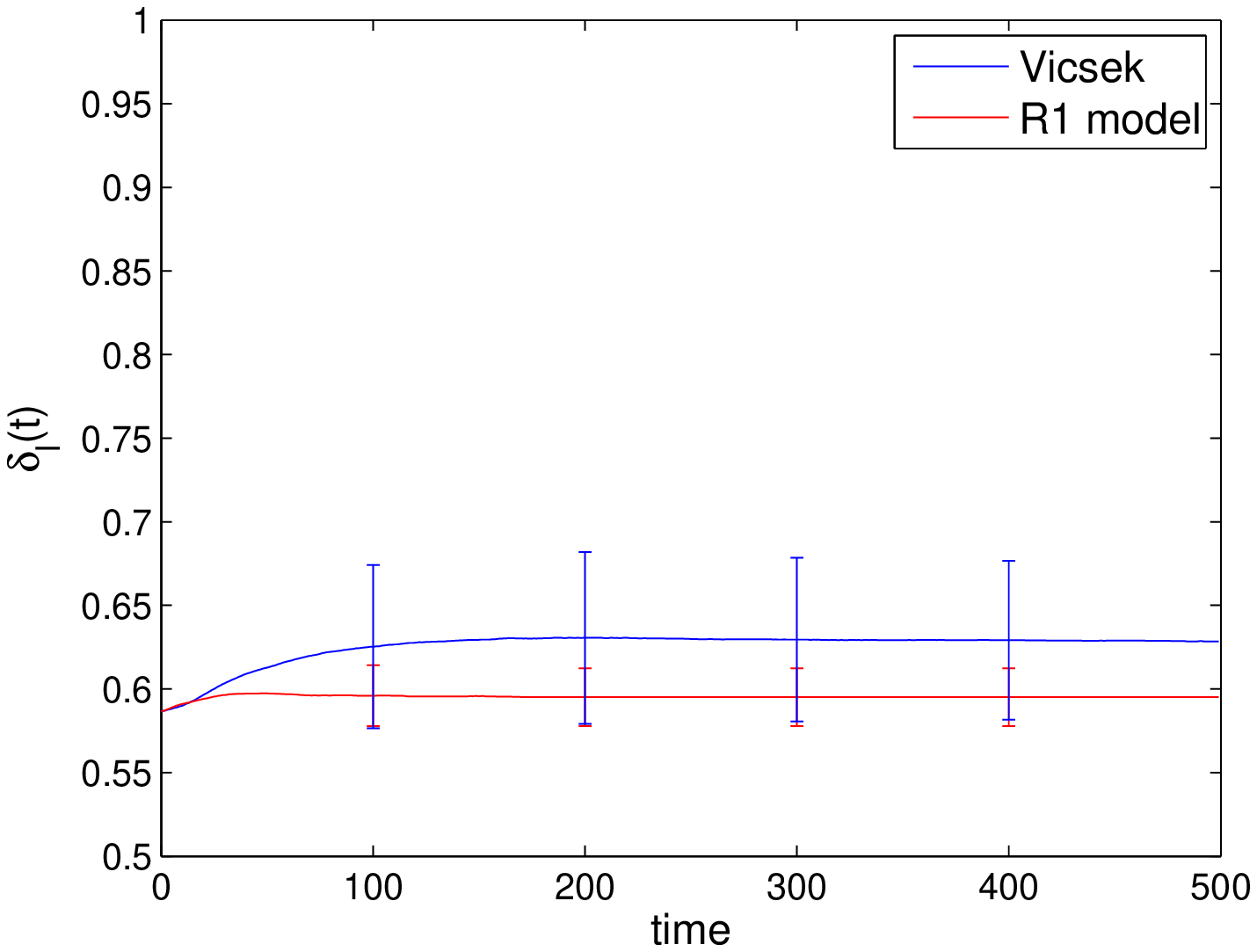} }
  \caption{{\bf Comparison of $\mathbf{\delta_l(t)}$ between the modified Vicsek model data and the ``best" R1 model.} Statistics were averaged over 10 simulations, with both models using the same initial conditions.}
  \label{fig:vml}
\end{figure}

\begin{figure}[!ht]
  \centering
  \subfloat[Snapshot at t=10]{\label{fig:vms1}\includegraphics[scale=0.50]{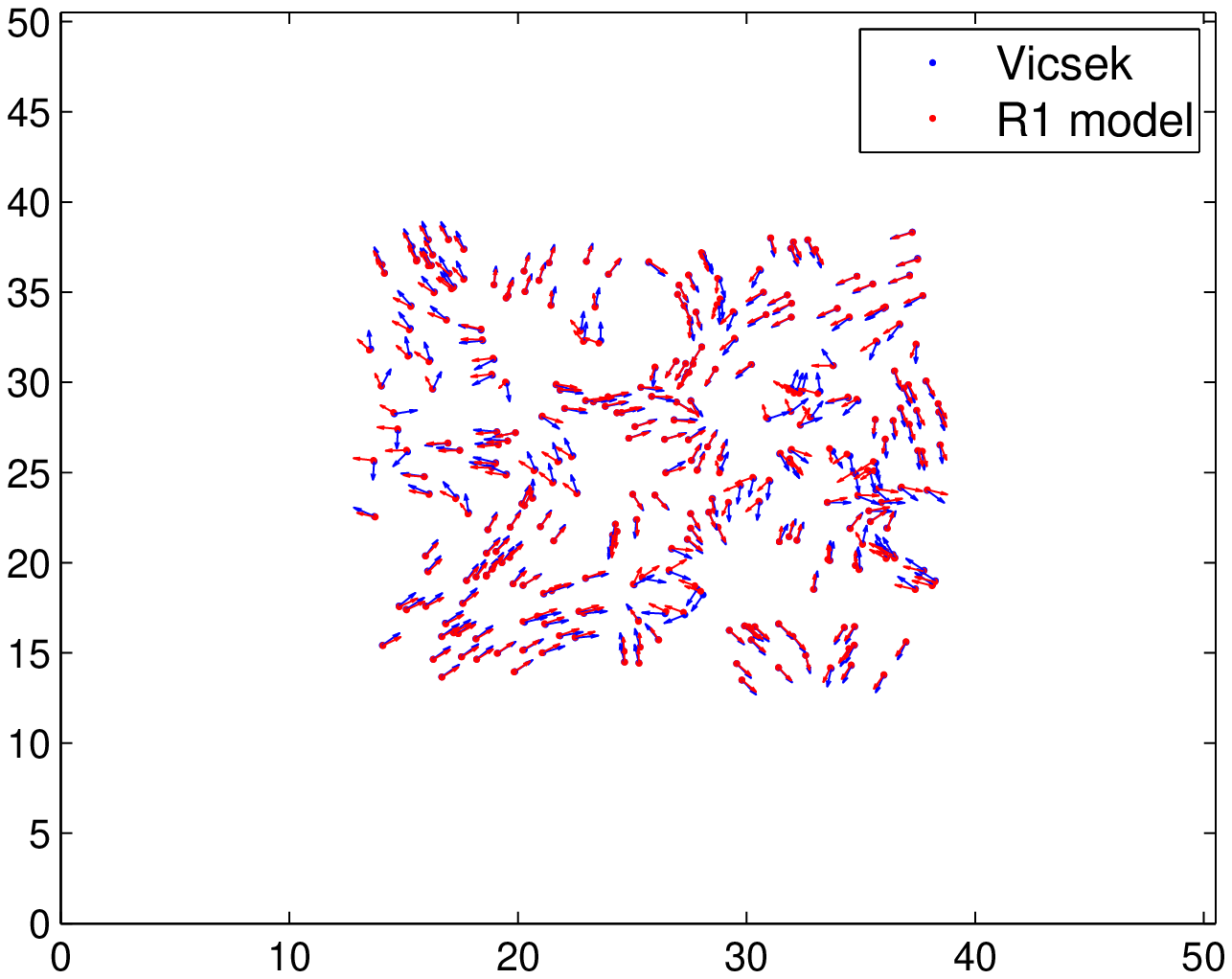}}                
  \subfloat[Snapshot at t=500]{\label{fig:vms2}\includegraphics[scale=0.50]{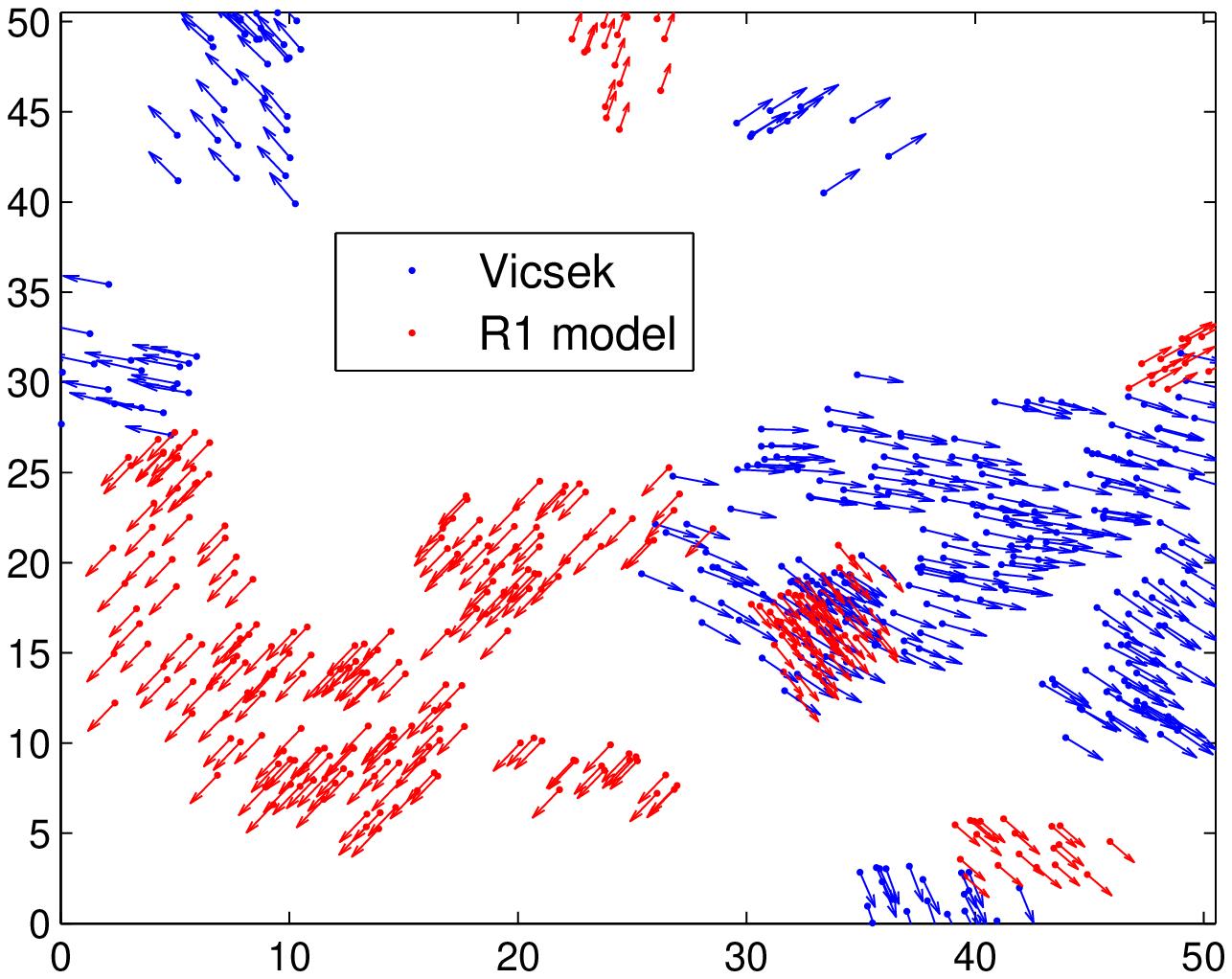} }
  \caption{{\bf Low density simulations of the modified Vicsek model and the ``best" R1 model.} The same initial conditions were used for both.}
  \label{fig:vmsl}
\end{figure} 

\begin{figure}[!ht]
  \centering
  \subfloat[Snapshot at t=25]{\label{fig:vms3}\includegraphics[scale=0.50]{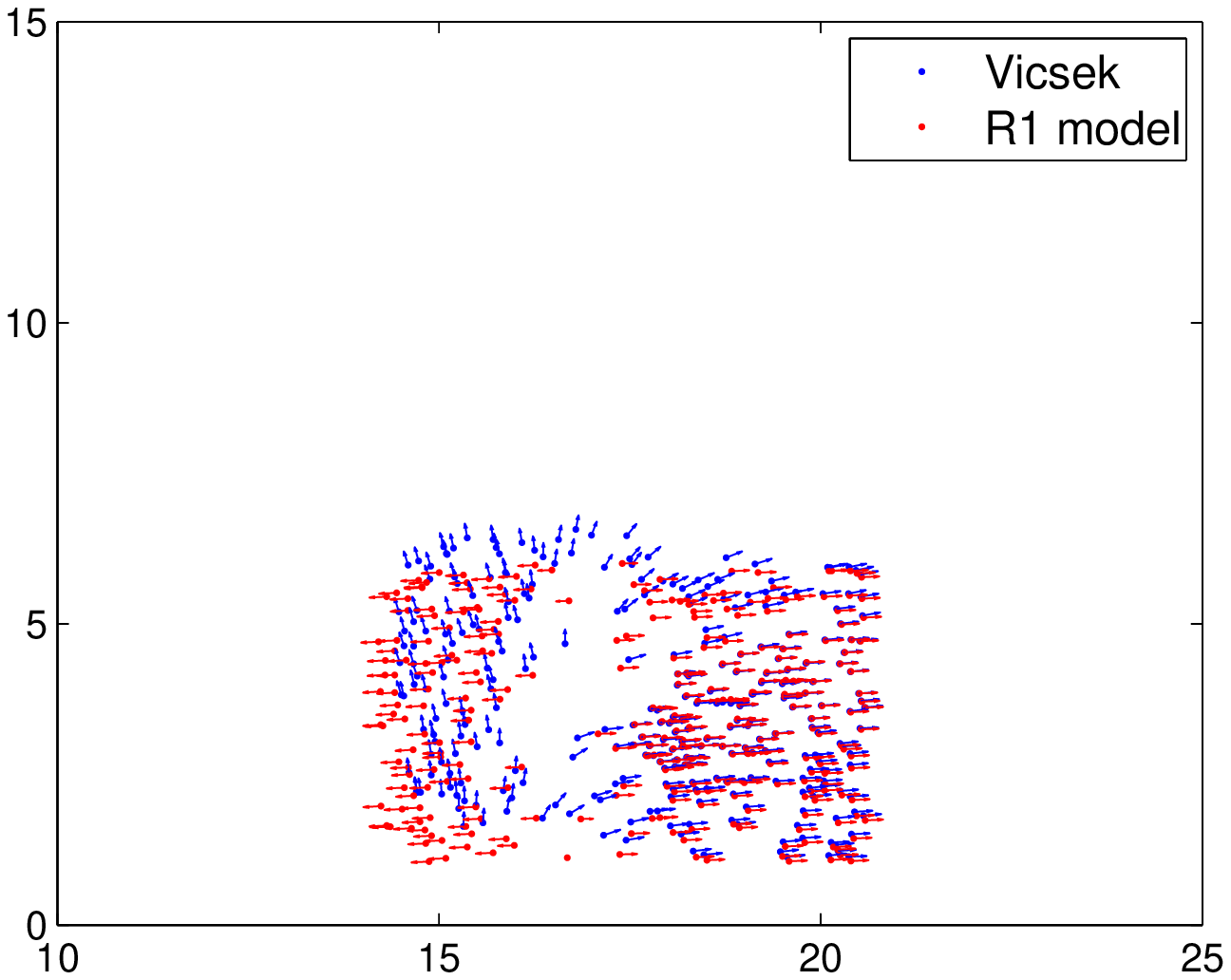}}                
  \subfloat[Snapshot at t=150]{\label{fig:vms4}\includegraphics[scale=0.50]{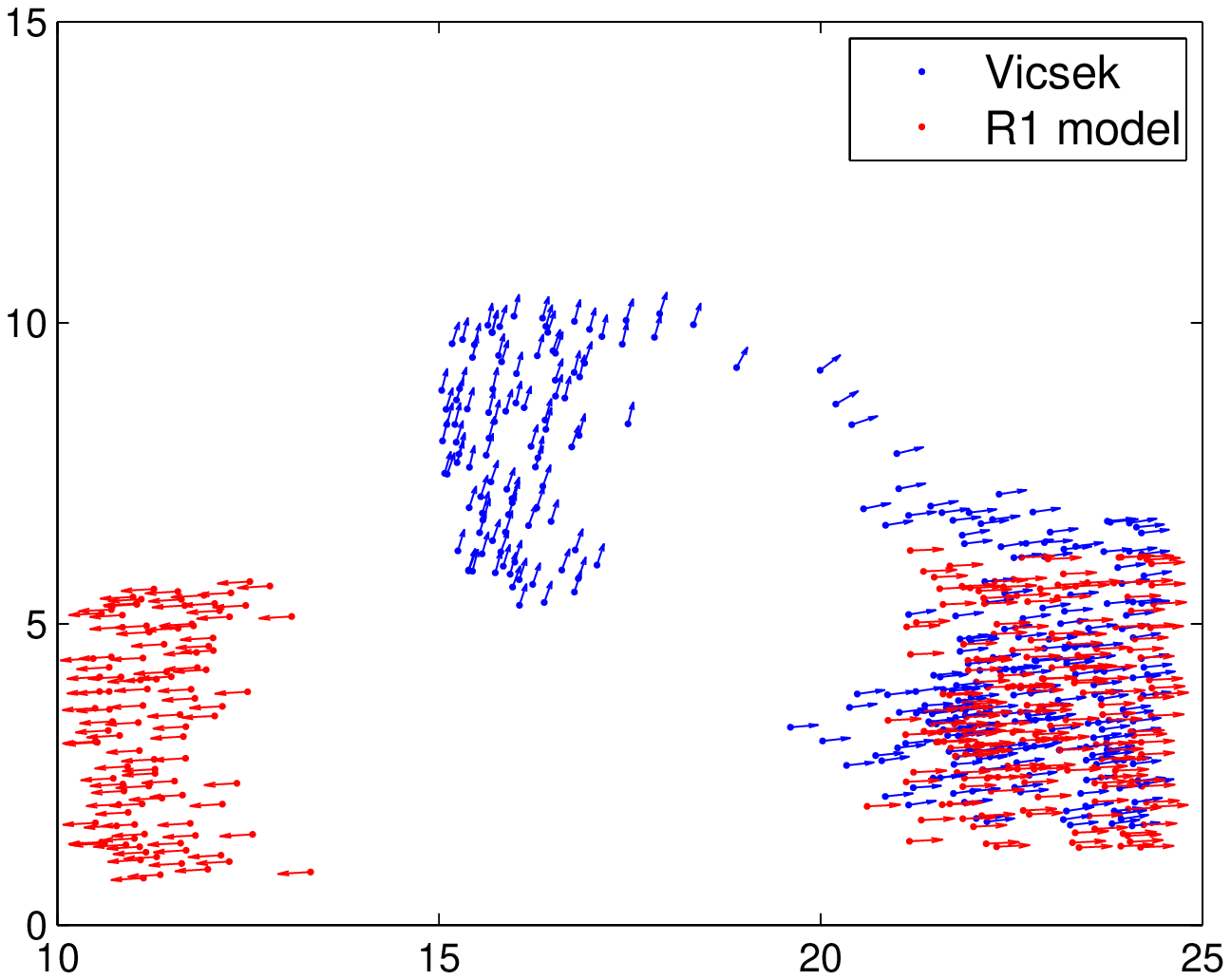} }
 \caption{{\bf High density simulations of the modified Vicsek model and the ``best" R1 model.} The same initial conditions were used for both.}
  \label{fig:vmsh}
\end{figure}

\begin{figure}[!ht]
  \centering
  \subfloat[Input data initial conditions]{\label{fig:a1}\includegraphics[scale=0.50]{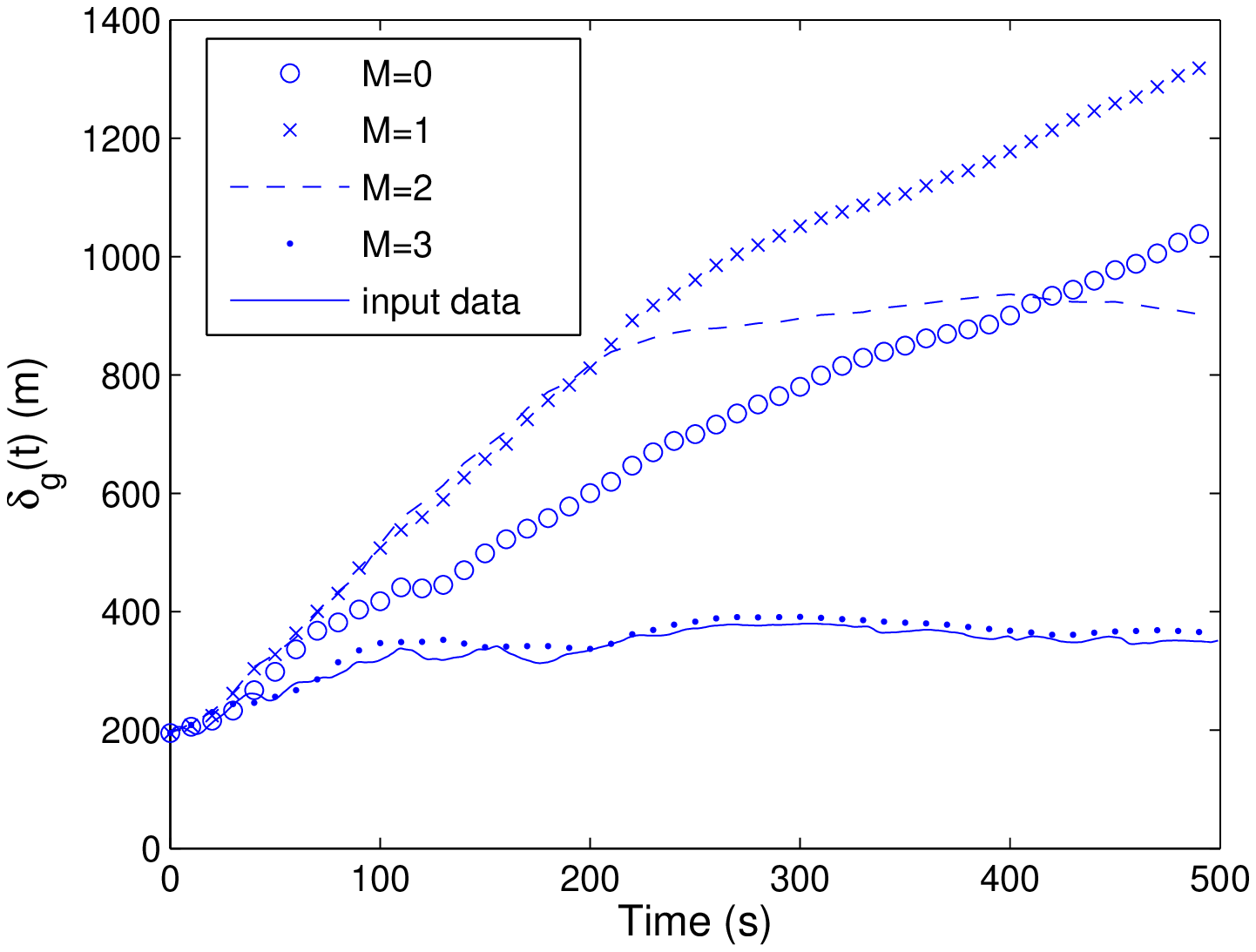}}                
  \subfloat[Random initial condtions]{\label{fig:a2}\includegraphics[scale=0.50]{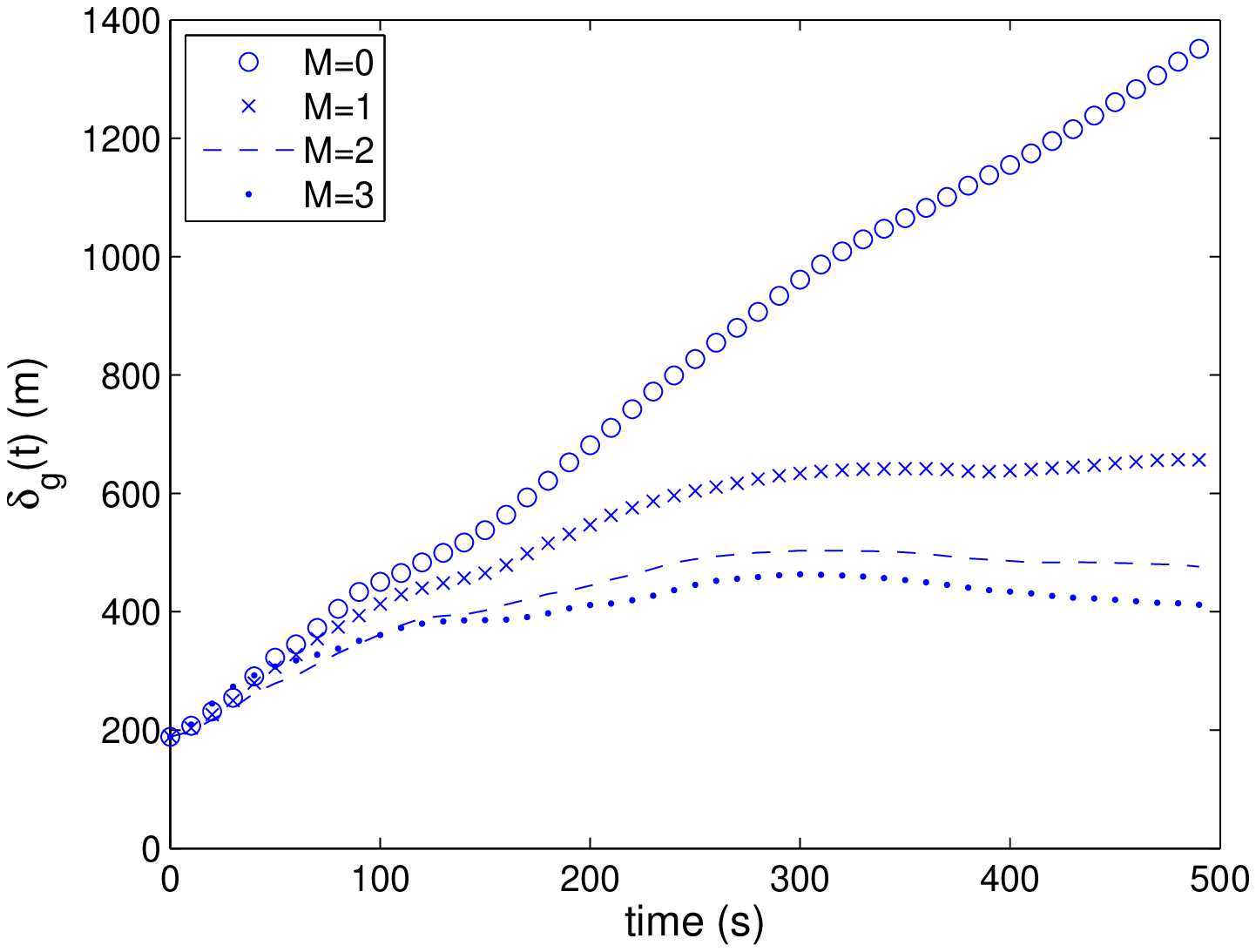} }
  \caption{{\bf Homing flight 1: flock separation for models with different interaction structure.} In (a), simulations consider initial conditions from the input data, while (b) averages over ten simulations of random initial conditions.}
  \label{fig:hf1}
\end{figure}

\begin{figure}[!ht]
  \centering
  \subfloat[Input data initial conditions]{\label{fig:a3}\includegraphics[scale=0.50]{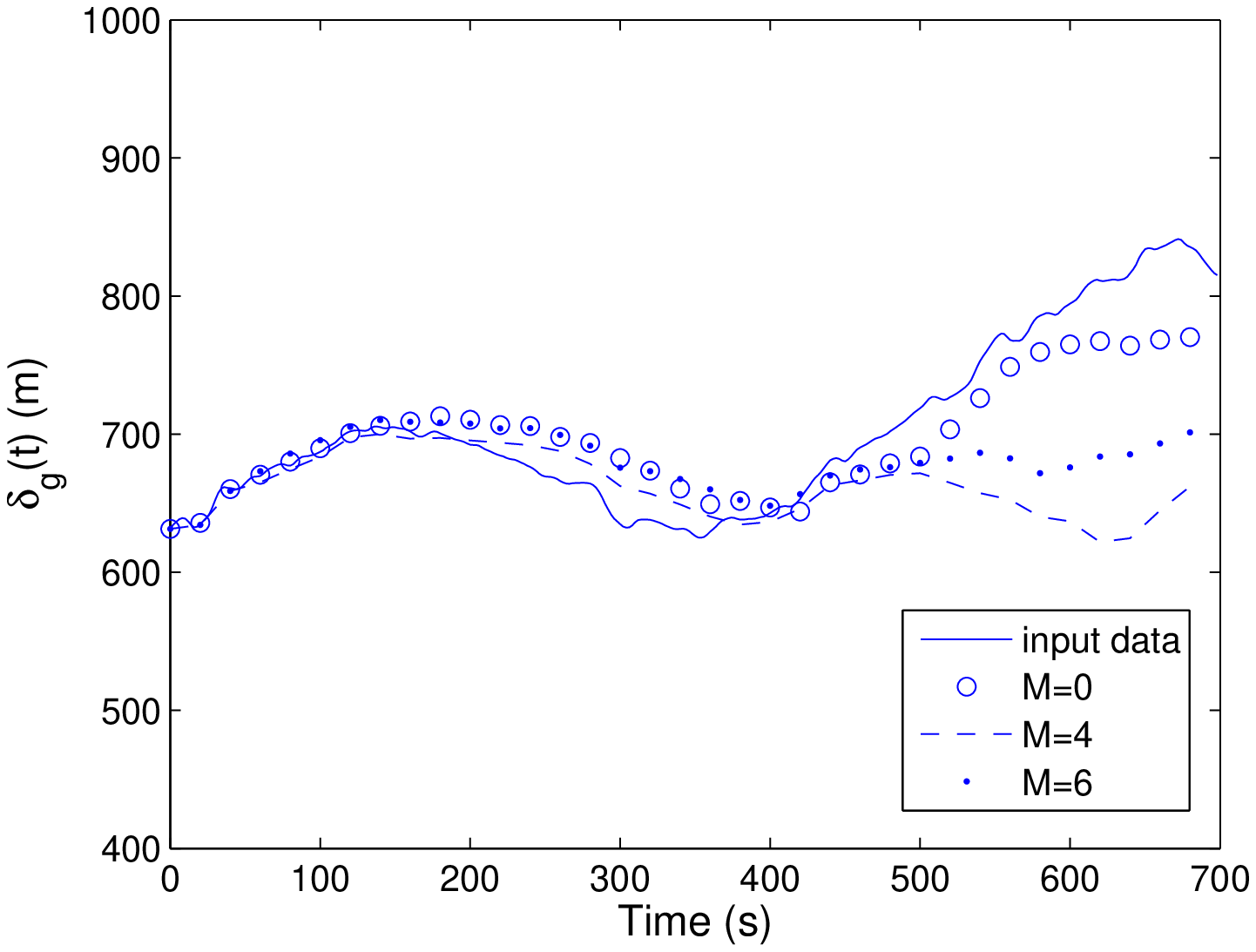}}                
  \subfloat[Random initial condtions]{\label{fig:a4}\includegraphics[scale=0.50]{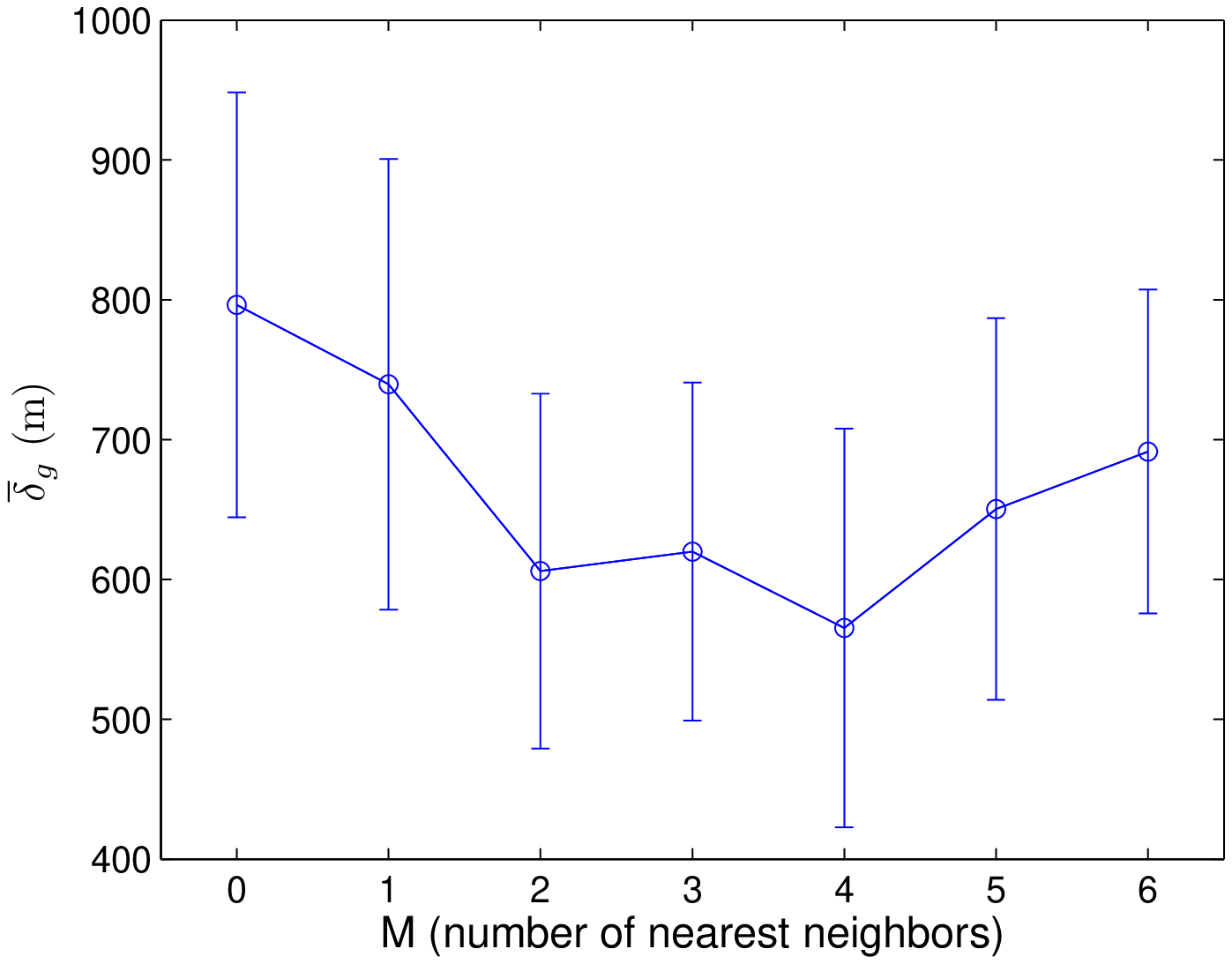} }
  \caption{{\bf Homing flight 2: flock separation for models with different interaction structure.} In (a), simulations consider initial conditions from the input data, while (b) averages over ten simulations of random initial conditions, and all time intervals.}
  \label{fig:hf2}
\end{figure}

\begin{figure}[!ht]
  \centering
  \subfloat[Input data initial conditions]{\label{fig:a5}\includegraphics[scale=0.50]{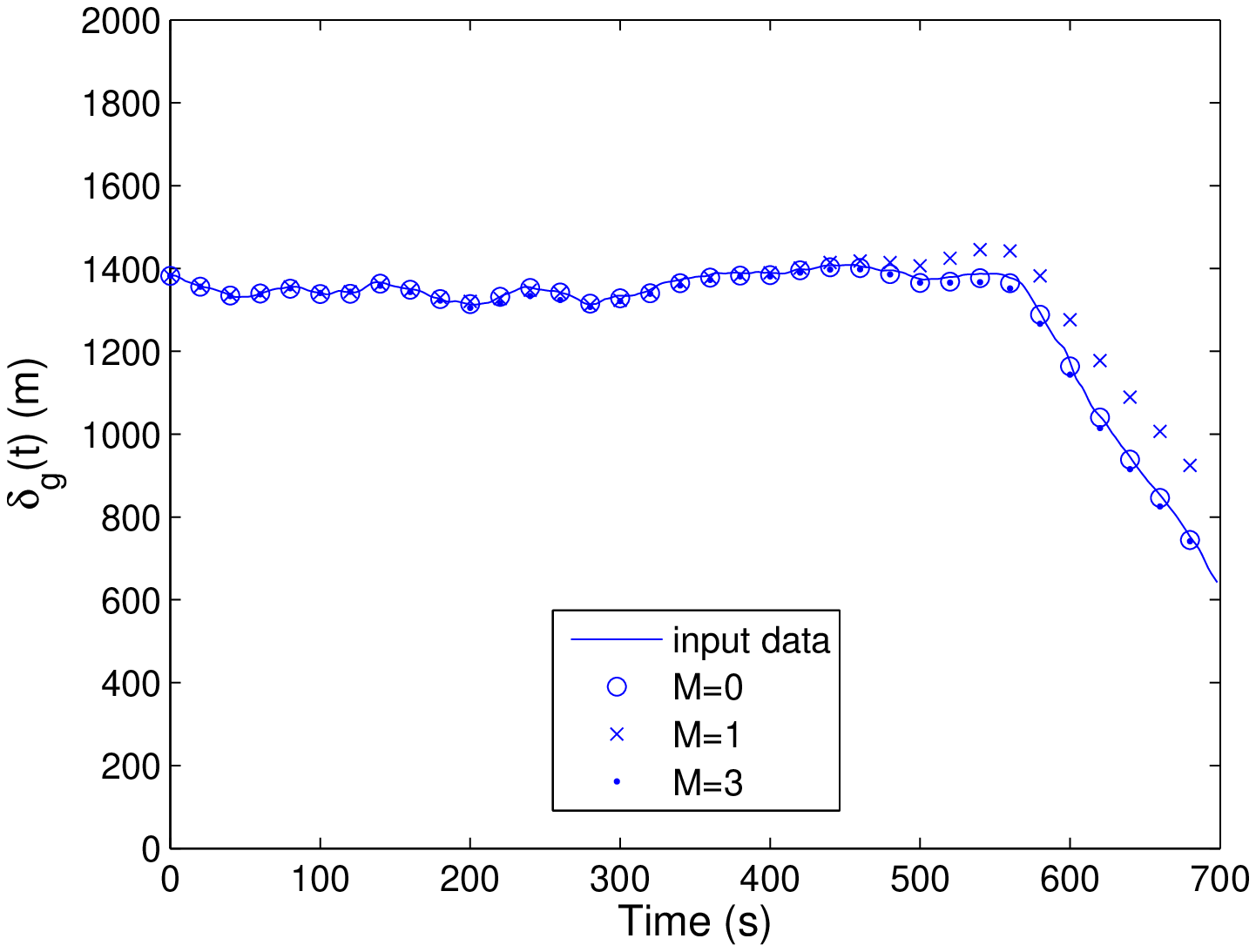}}                
  \subfloat[Random initial conditions]{\label{fig:a6}\includegraphics[scale=0.50]{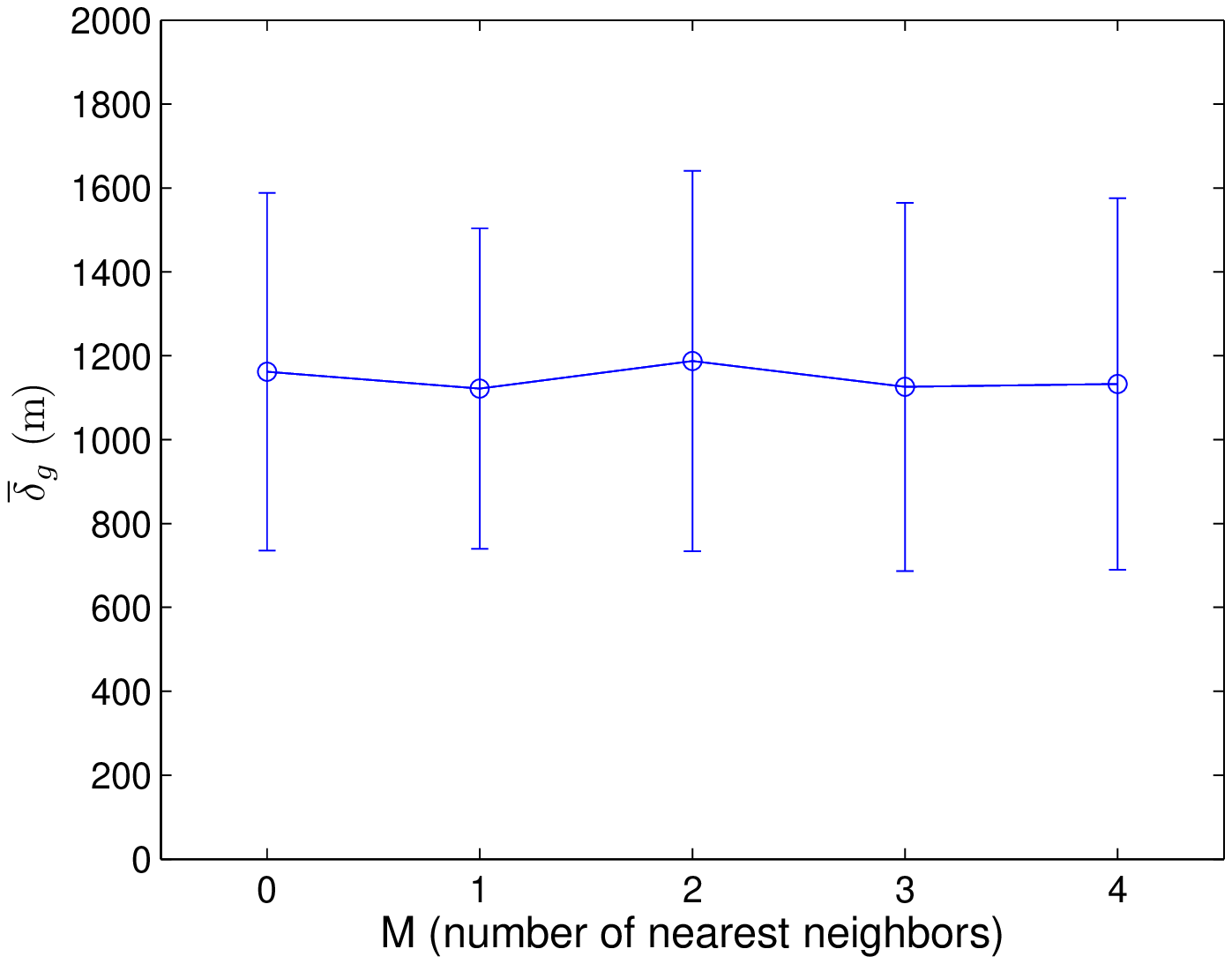} }
  \caption{{\bf Homing flight 3: Flock separation for models with different interaction structure.} In (a), simulations consider initial conditions from the input data, while (b) averages over ten simulations of random initial conditions, and all time intervals.}
  \label{fig:hf3}
\end{figure}

\begin{figure}[!ht]
  \centering
  \subfloat[Input data initial conditions]{\label{fig:a7}\includegraphics[scale=0.50]{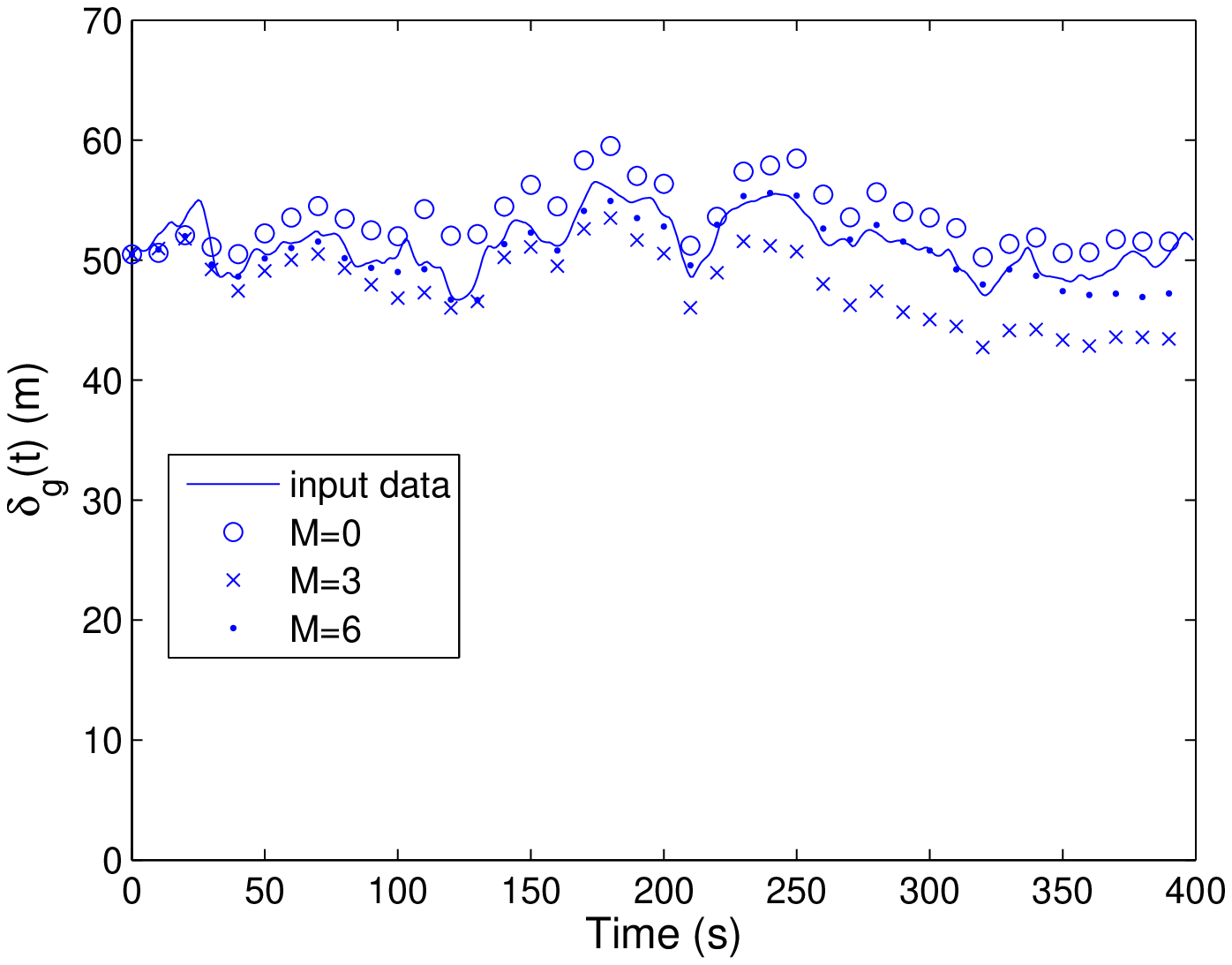}}                
  \subfloat[Random initial condtions]{\label{fig:a8}\includegraphics[scale=0.50]{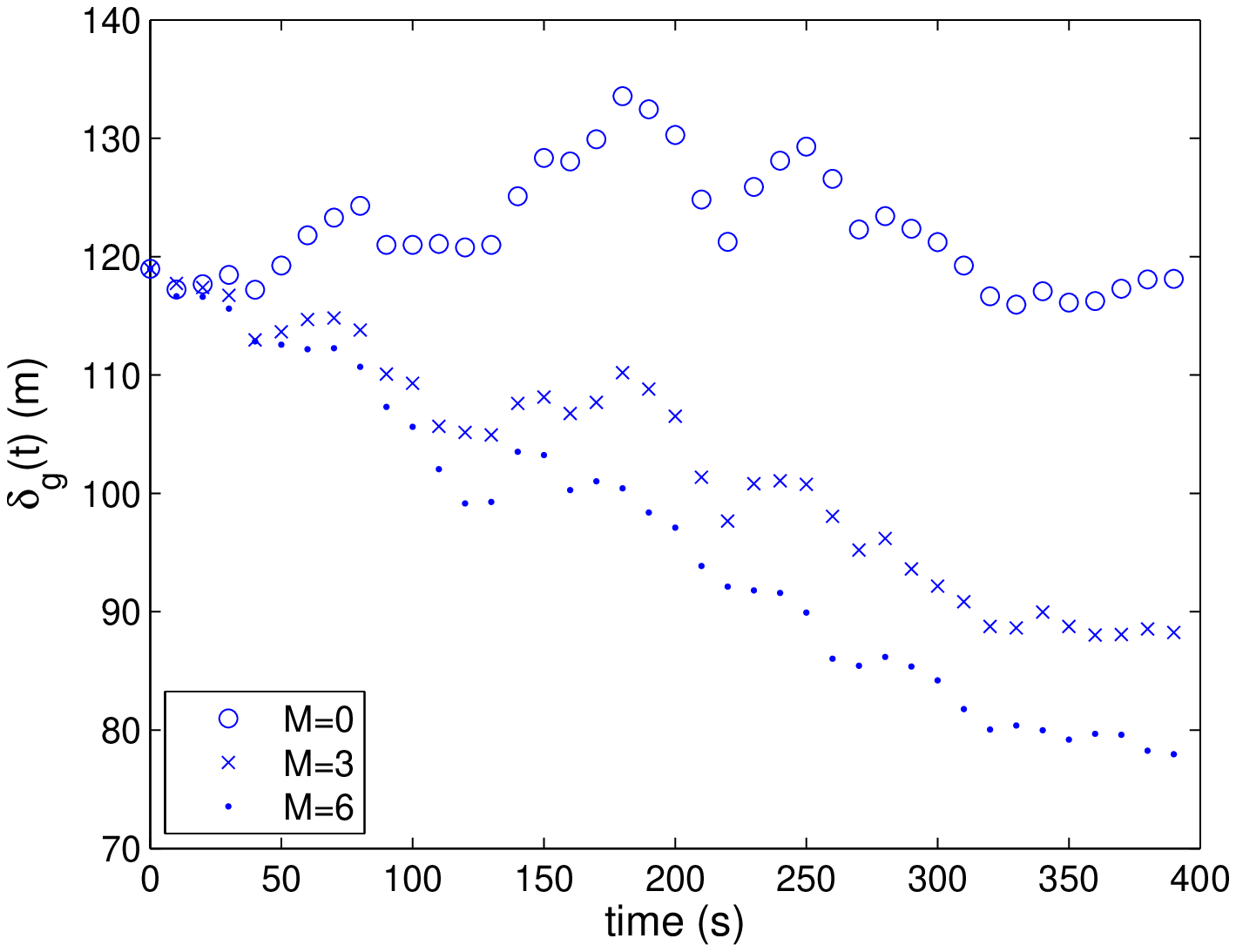} }
  \caption{{\bf Homing flight 4: Flock separation for models with different interaction structure.} In (a), simulations consider initial conditions from the input data, while (b) averages over ten simulations of random initial conditions.}
  \label{fig:hf4}
\end{figure}

\begin{figure}[!ht]
  \centering
  \subfloat[Global separation]{\label{fig:m1}\includegraphics[scale=0.50]{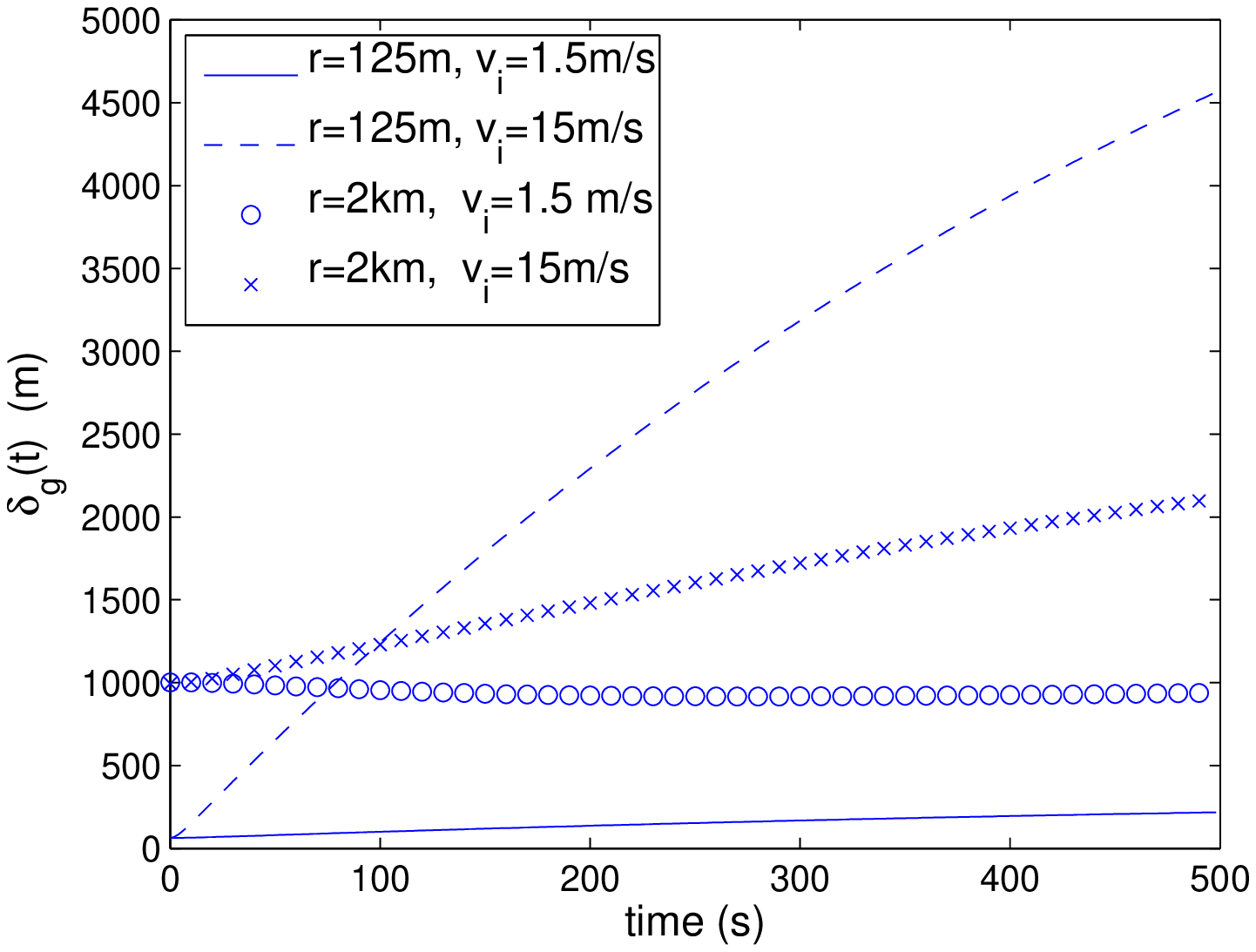}}                
  \subfloat[Local separation]{\label{fig:m2}\includegraphics[scale=0.50]{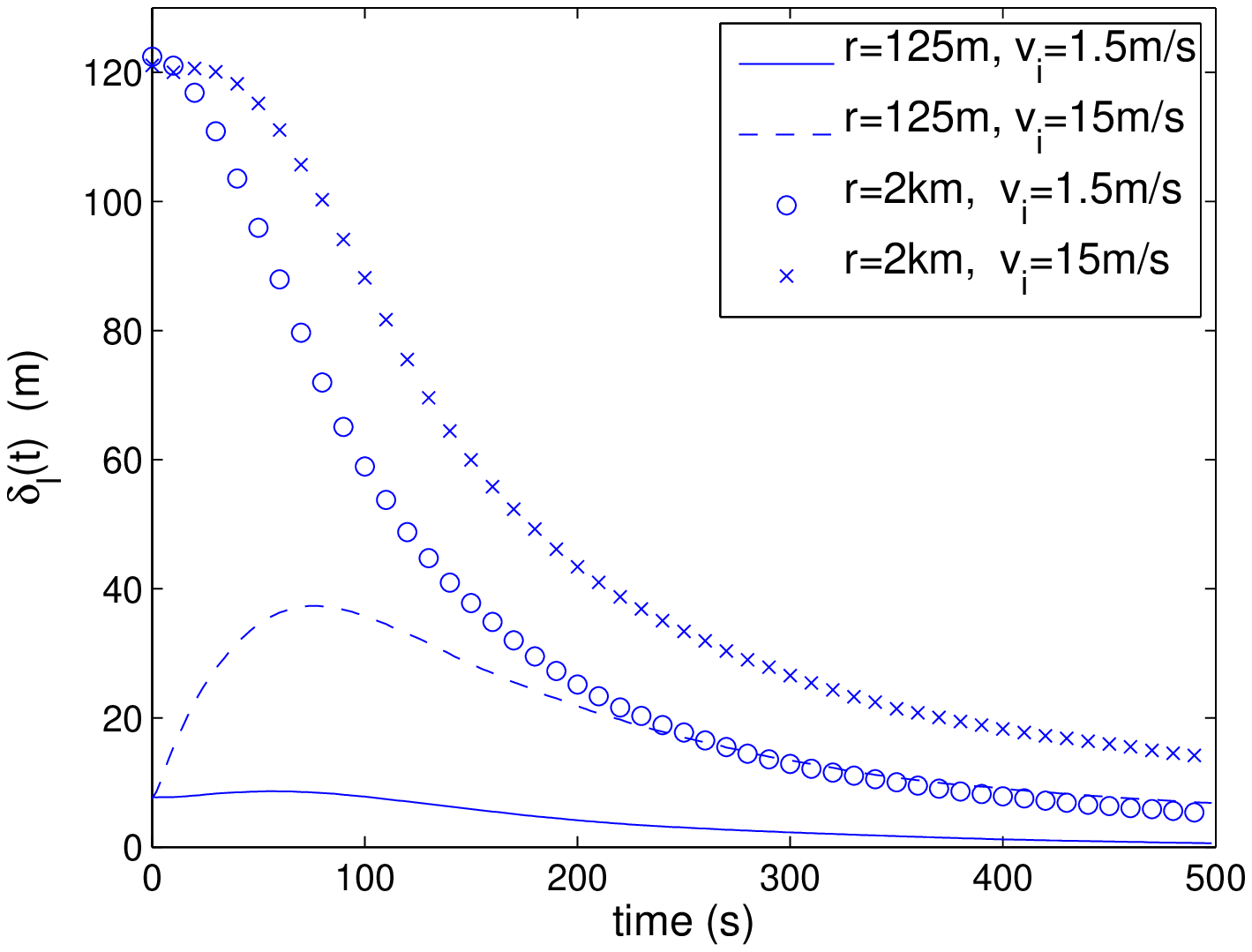} }
  \caption{{\bf Comparison of separation measures between extreme cases of initial conditions for the R2 model.} Statistics were averaged over 10 simulations for each case.}
  \label{fig:msep}
\end{figure}

\begin{figure}[!ht]
  \centering
  \subfloat[Snapshot after 100 s]{\label{fig:ms1}\includegraphics[scale=0.50]{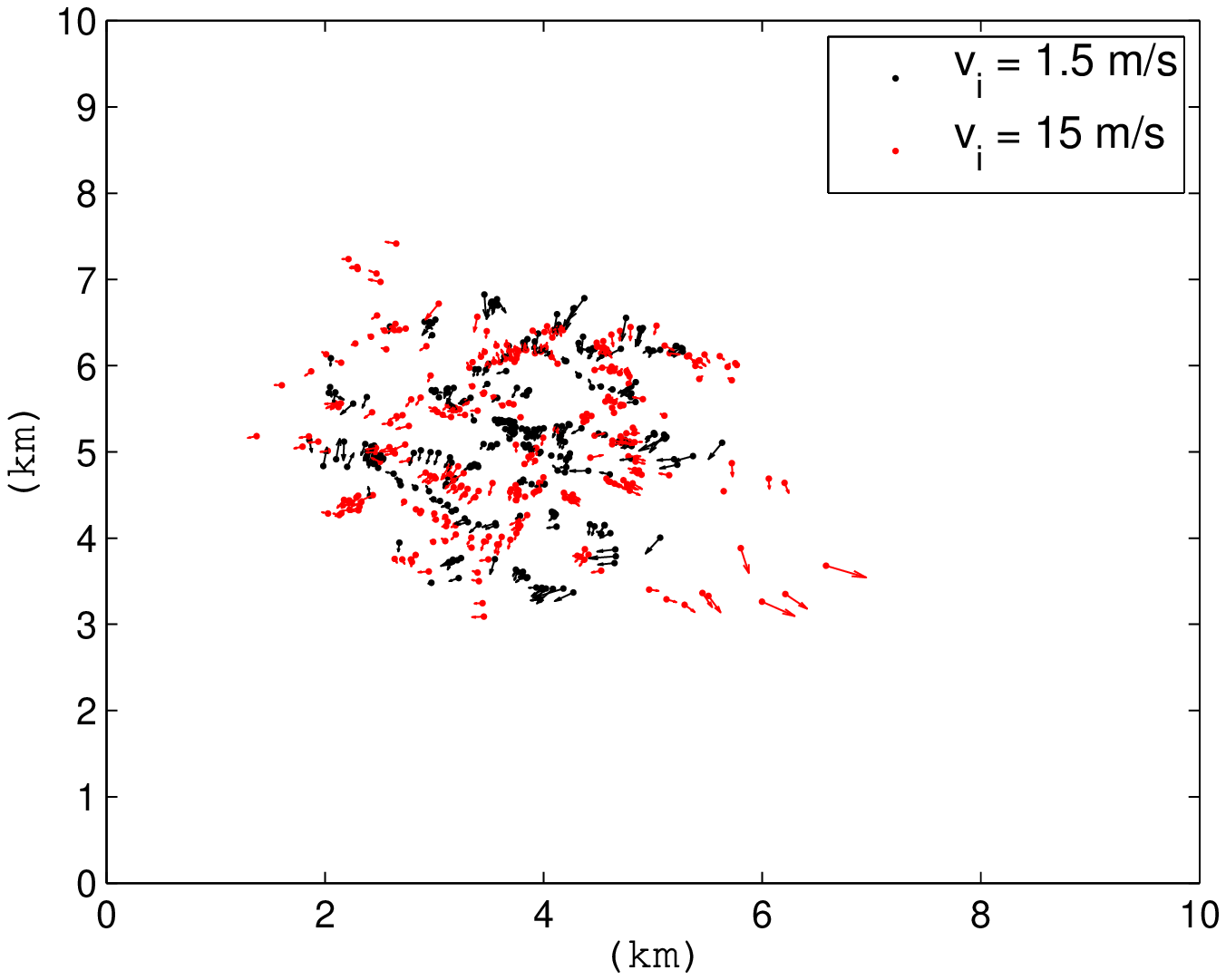}}                
  \subfloat[Snapshot after 500 s]{\label{fig:ms2}\includegraphics[scale=0.50]{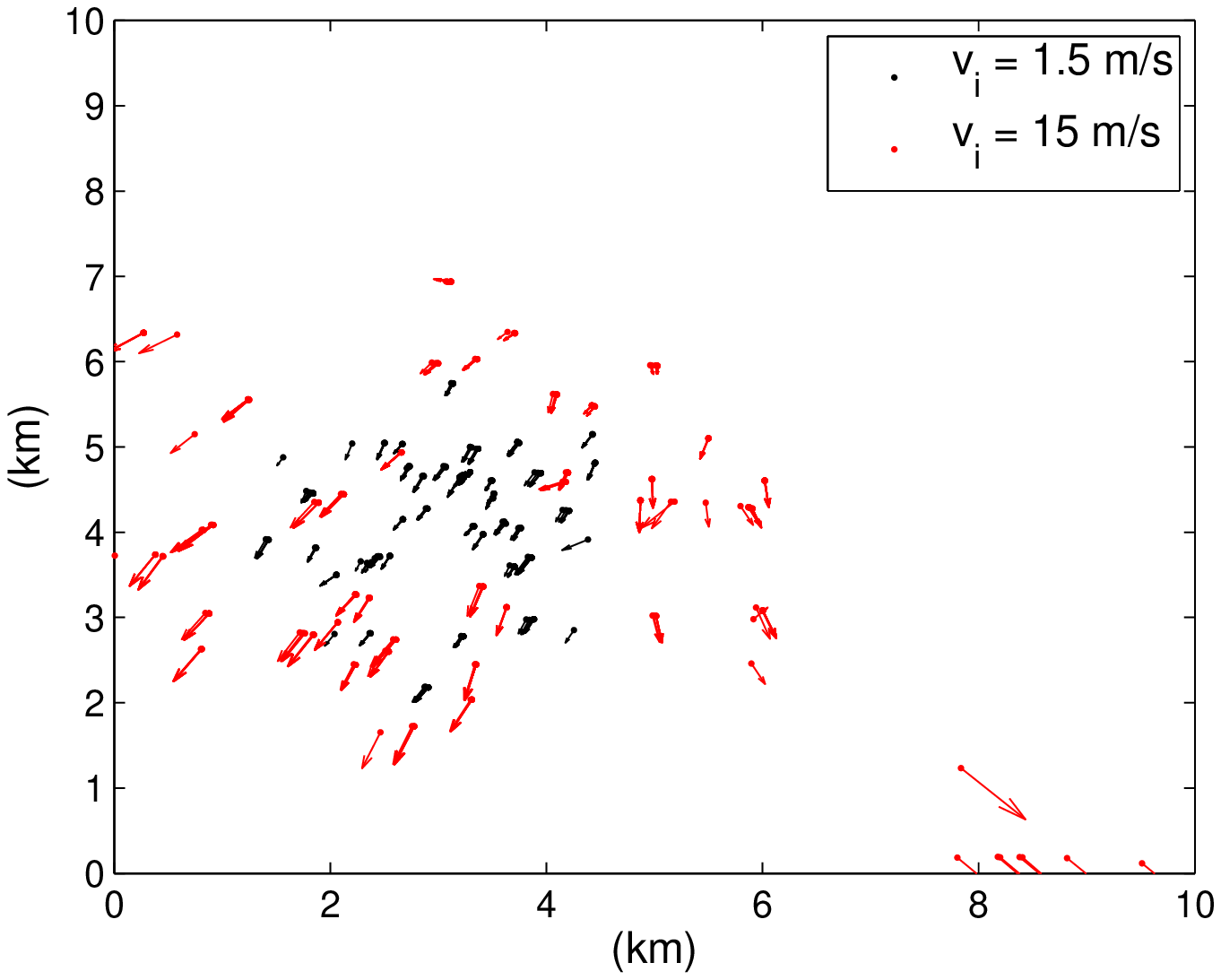} }
  \caption{{\bf Simulations of low density instances ($\mathbf{r = 2 km}$) of the R2 model.} Low and high initial speeds are considered.}
  \label{fig:msl}
\end{figure}

\begin{figure}[!ht]
  \centering
  \subfloat[Snapshot after 100 s]{\label{fig:ms3}\includegraphics[scale=0.50]{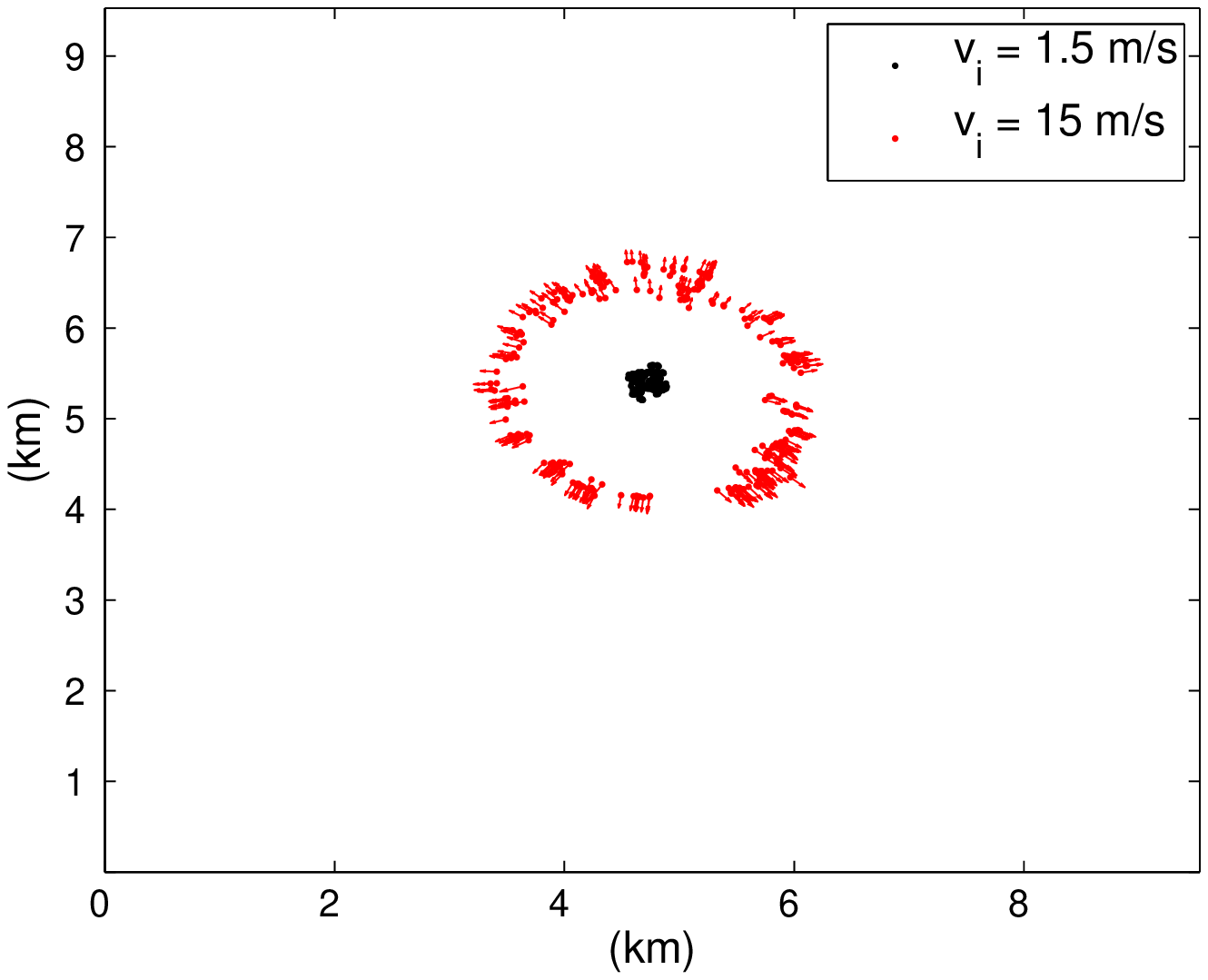}}               
  \subfloat[Snapshot after 500 s]{\label{fig:ms4}\includegraphics[scale=0.50]{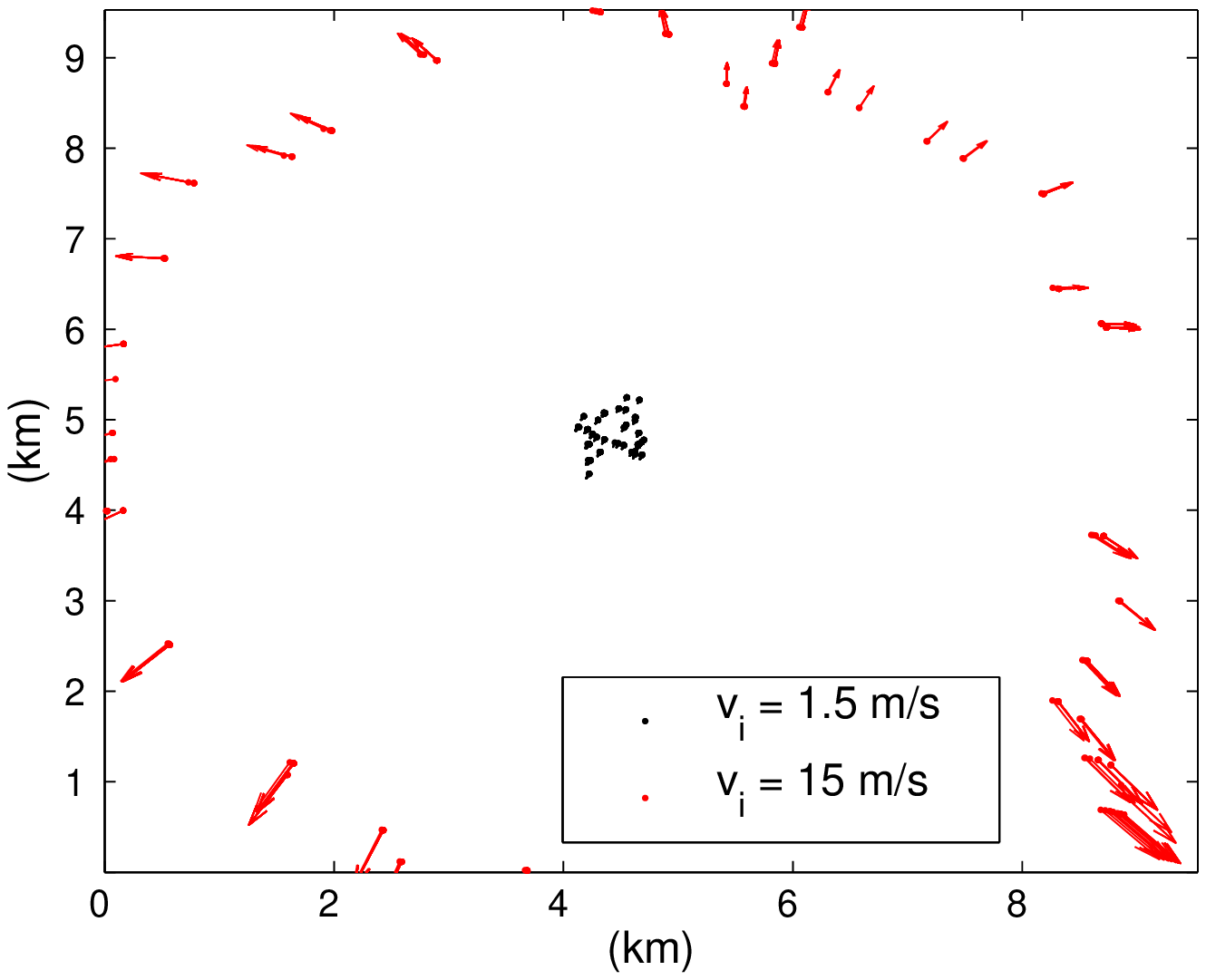} }
 \caption{{\bf Simulations of high density instances ($\mathbf{r = 125 m}$) of the R2 model.} Low and high initial speeds are considered.}
  \label{fig:msh}
\end{figure}

\begin{figure}[!ht]
  \centering
  \subfloat[Averaged global separation $\overline{\delta}_g$]{\label{fig:m3}\includegraphics[scale=0.50]{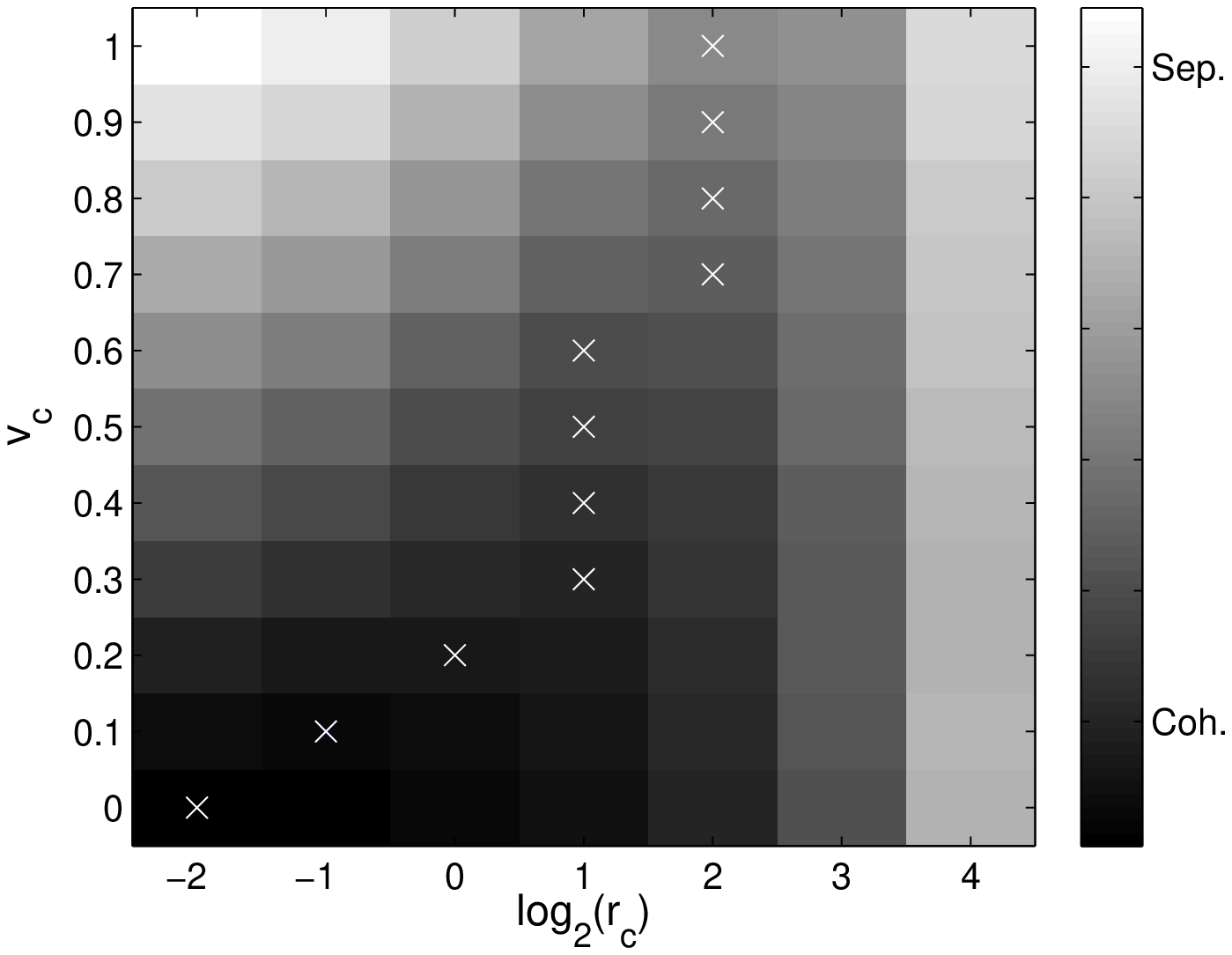}}                
  \subfloat[Averaged local separation $\overline{\delta}_l$]{\label{fig:m4}\includegraphics[scale=0.50]{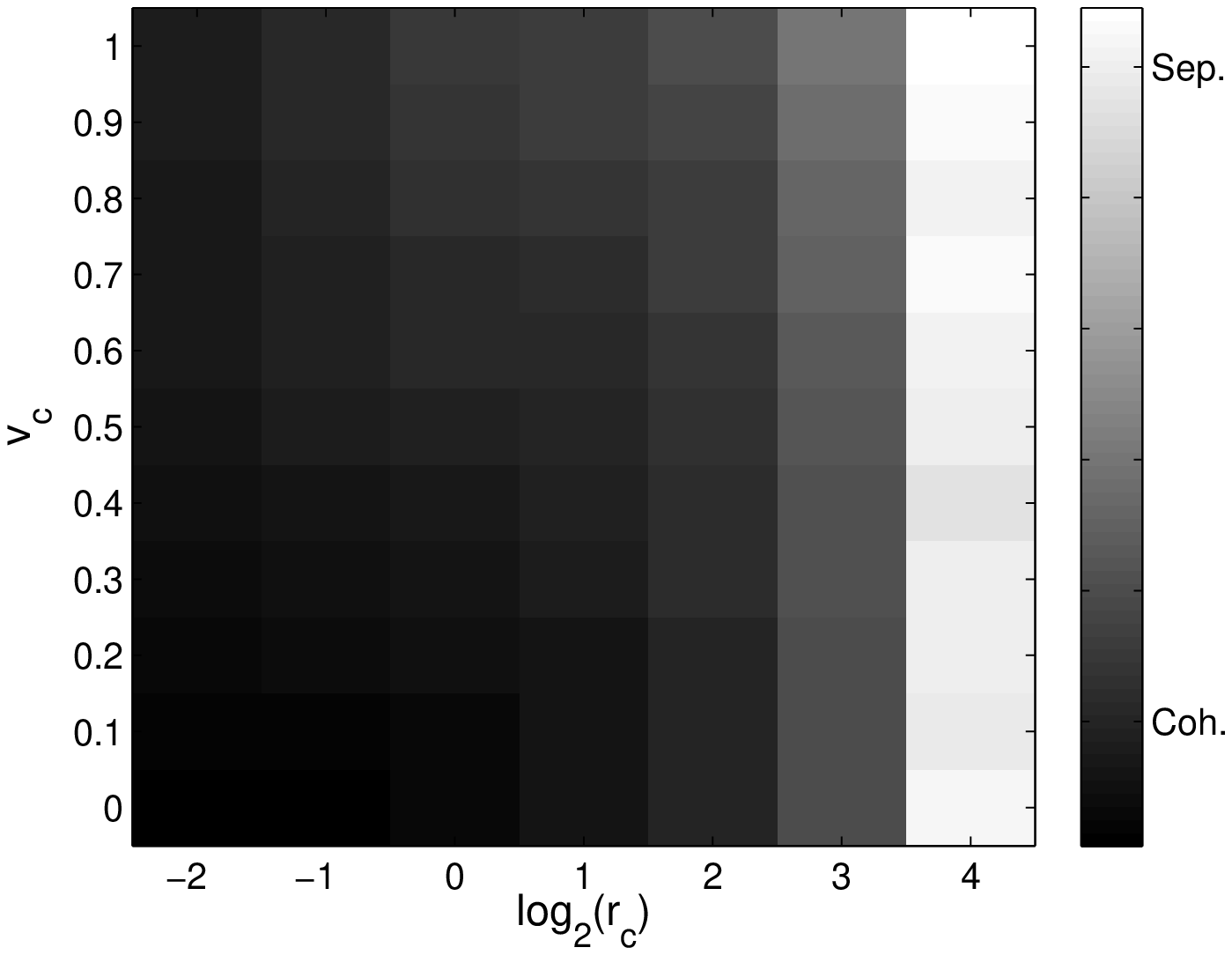} }
  \caption{{\bf Comparison of averaged separation measure $\mathbf{\overline{\delta}}$ for simulations of R2 model} Initial speeds and population radius (densities) for the R2 model were varied. Critical density values (highest cohesion) for each speed are marked in (a). Statistics were averaged over 10 simulations for each parameter case.}
  \label{fig:mcolor}
\end{figure}

\begin{figure}[!ht]
  \centering
  \subfloat[Change in separation to nearest neighbors]{\label{fig:m5}\includegraphics[scale=0.50]{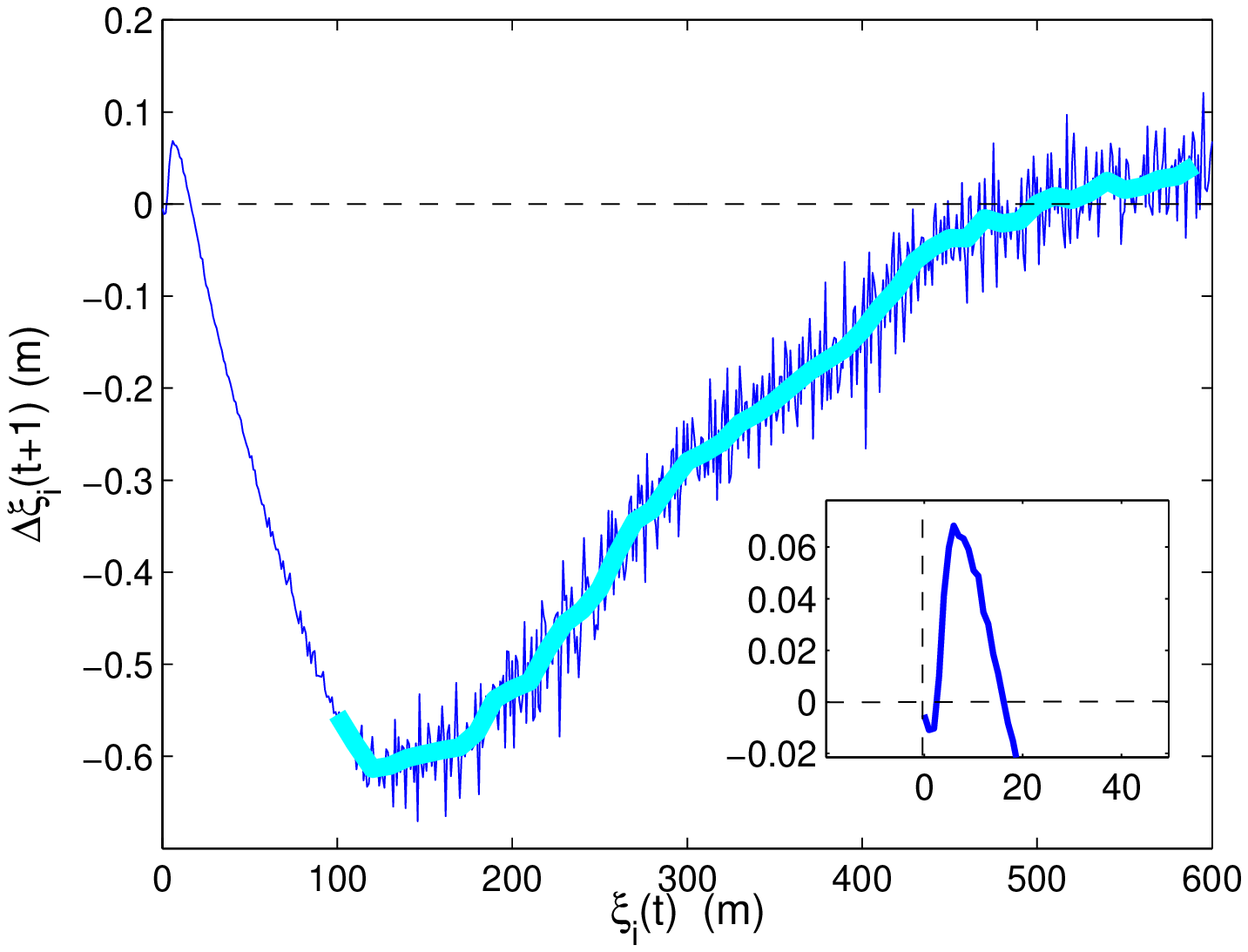}}                
  \subfloat[Speed of a single particle]{\label{fig:m6}\includegraphics[scale=0.50]{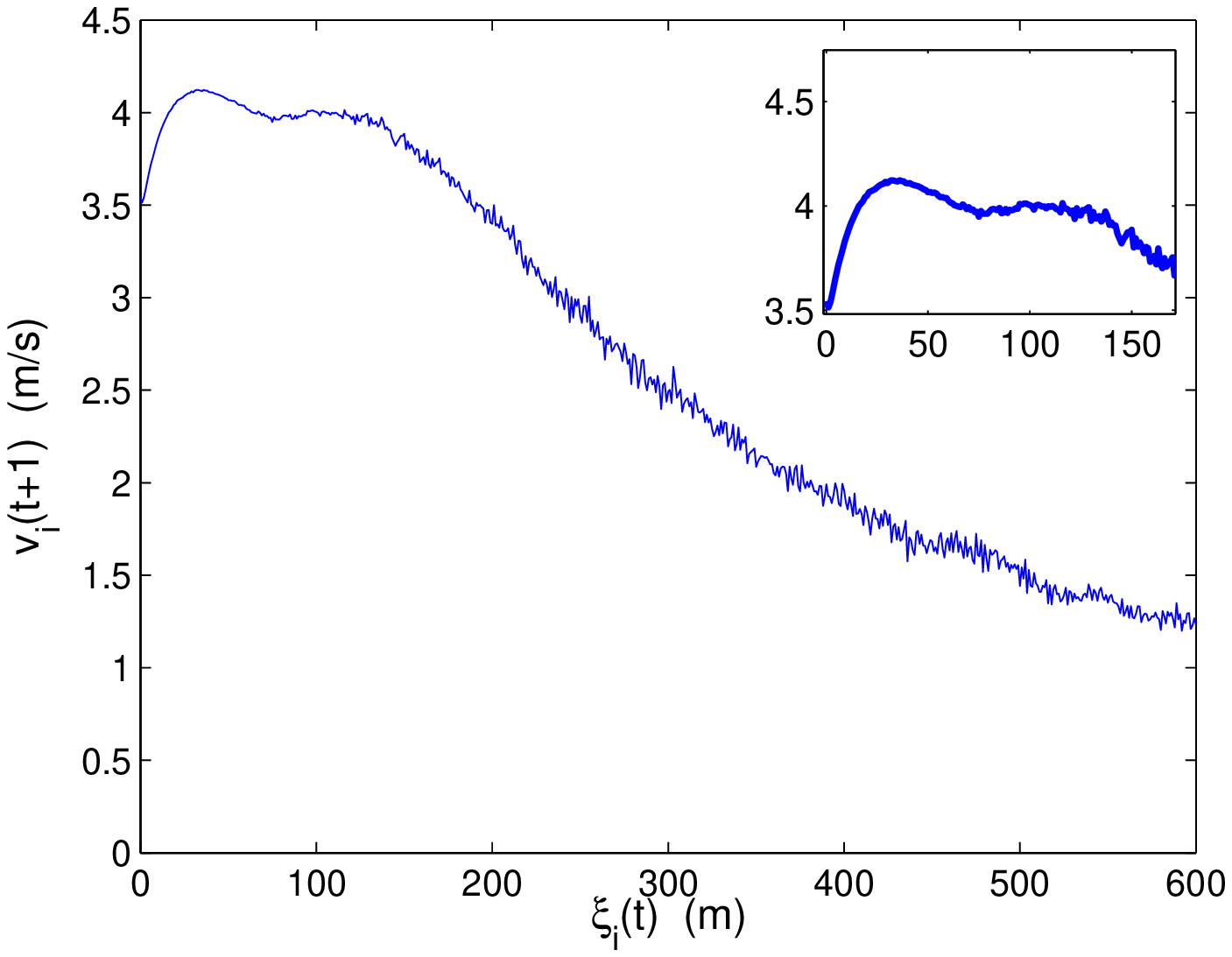} }
  \caption{{\bf Change in separation symbolizing attraction(-)/repulsion(+),  and speed distributions estimated from the simulations.} In (a) and (b) we have the changes in neighbor separation $\Delta \xi_i$ and speed \emph{v} respectively at the next time interval $t+1$, as a function of $\xi_i$ at time t. The R2 model was built upon two-second sampling, and thus each time interval update $t+1$ is after two seconds.}
  \label{fig:attract}
\end{figure}



\end{document}